\documentclass[aps, prd, 11pt, onecolumn, tightenlines, secnumarabic, superscriptaddress, preprintnumbers, nofootinbib]{revtex4-2}

\pdfoutput=1
\usepackage[utf8]{inputenc}
\usepackage[T1]{fontenc}
\usepackage{amsmath, amssymb, bm, slashed}
\usepackage{booktabs, tabularx}
\usepackage[labelformat=empty]{subcaption}
\usepackage{hyperref, cleveref}
\usepackage{graphicx}
\usepackage[normalem]{ulem}

\usepackage{feynmp-auto}

\begin{document}
	
%============================================================================

\title{Has the Origin of the Third-Family Fermion Masses been Determined?}
	
\author{Michael J.\ Baker}
\email{michael.baker@unimelb.edu.au}
\affiliation{ARC Centre of Excellence for Dark Matter Particle Physics, 
School of Physics, The University of Melbourne, Victoria 3010, Australia}
	
\author{Peter Cox}
\email{peter.cox@unimelb.edu.au}
\affiliation{ARC Centre of Excellence for Dark Matter Particle Physics, 
School of Physics, The University of Melbourne, Victoria 3010, Australia}
	
\author{Raymond R. Volkas}
\email{raymondv@unimelb.edu.au}
\affiliation{ARC Centre of Excellence for Dark Matter Particle Physics, 
School of Physics, The University of Melbourne, Victoria 3010, Australia}

%============================================================================
	
\begin{abstract}
Precision measurements of the Higgs couplings are, for the first time, directly probing the mechanism of fermion mass generation. The purpose of this work is to determine to what extent these measurements can distinguish between the tree-level mechanism of the Standard Model and the theoretically motivated alternative of radiative mass generation. Focusing on the third-family, we classify the minimal one-loop models and find that they fall into two general classes. By exploring several benchmark models in detail, we demonstrate that a radiative origin for the tau-lepton and bottom-quark masses is consistent with current observations. While future colliders will not be able to rule out a radiative origin, they can probe interesting regions of parameter space.  
\end{abstract}

%============================================================================
\maketitle
%============================================================================

%=============================================================================
\section{Introduction}
\label{sec:new-introduction}
%=============================================================================

The Standard Model (SM) predicts that the Yukawa coupling constants $y_f$ of the quarks and the charged leptons to the physical Higgs boson $h$ are equal to the respective fermion masses $m_f$ divided by the Higgs vacuum expectation value (vev),
\begin{equation}
    y_f = \sqrt{2}\, \frac{m_f}{v}\,,
    \label{eq:lambda-m-treelevel}
\end{equation}
where the vev is $v/\sqrt{2}$. The purpose of this paper is to analyse if this minimal fermion mass generation mechanism has either already been experimentally established for the $b$-quark and the $\tau$-lepton, or could be established in the foreseeable future. This requires measuring $y_f$ and $m_f$ independently and determining whether these measurements can distinguish between the SM, with its prediction in \cref{eq:lambda-m-treelevel}, and alternative hypotheses.  

In our analysis, we contrast the minimal SM mechanism with the logical alternative of radiative mass generation. Radiative scenarios are motivated by the fact that the charged fermion masses, save that of the top quark, are measured to be orders of magnitude lower than the electroweak scale, set by $v \simeq 246$ GeV. While it is technically natural for the Yukawa couplings in \cref{eq:lambda-m-treelevel} to be much smaller than one, having the fermion masses generated through radiative effects, perhaps at various loop levels, is a theoretically attractive option. This has the potential to explain the hierarchies amongst the masses, a fundamental aspect of the flavour puzzle. As such, radiative mass generation has been studied extensively (see~\cite{
10.1016/0550-3213(81)90337-0,10.1142/S0217732389002239,10.1103/PhysRevLett.64.2747,10.1103/PhysRevLett.64.2866,10.1103/PhysRevD.41.1630,10.1103/PhysRevLett.66.556,10.1016/S0550-3213(99)00328-4,10.1088/1126-6708/2008/08/100,10.1103/PhysRevLett.112.091801,Fraser:2014ija,10.1140/epjc/s10052-016-4351-y,10.1007/JHEP02(2017)125,10.1007/JHEP06(2020)043,Morais:2020ypd} for works which treat the $b$-quark and $\tau$-lepton masses radiatively, and references therein for the wider literature).  Now that the Higgs boson has been discovered and its couplings are being measured it is pertinent to revisit this idea and determine to what extent it is still viable.

Currently, the LHC experiments have measured the couplings of the Higgs boson to the top quark, bottom quark, tau lepton, and very recently the muon. There is an upper bound on the coupling to the charm quark. Our focus in this paper will therefore be on the third-family charged fermions, with the muon treated in a companion paper~\cite{2103.13401}.

The coupling to the top quark has been measured through $t \overline{t} H$ production, with the signal strength relative to the SM expectation found to be $1.28 \pm 0.20$~\cite{Zyla:2020zbs}, consistent with the SM at just over $1\sigma$.  Since the top quark mass is near the electroweak scale, it is extremely unlikely to have a radiative origin. We thus assume in this paper that the top quark couples to the Higgs boson in exactly the same way and with the same strength as in the SM.  The signal strengths for Higgs production and subsequent decay to $b\overline{b}$ and $\tau^+ \tau^-$ are $1.04 \pm 0.13 $ and $1.15^{+0.16}_{-0.15}$, respectively~\cite{Zyla:2020zbs}. These results are completely consistent with the SM and thus also with \cref{eq:lambda-m-treelevel}. However, this should not be used to argue that the SM tree-level origin of the $b$-quark and $\tau$-lepton masses has been established, as the experimental results are also consistent with radiative generation, as we show in this paper. 

As more data are collected at the LHC, especially after the high-luminosity upgrade, and with the prospect of Higgs factories such as a 250 GeV ILC, the FCC-ee and CEPC being constructed, we expect the precision of the $b$ and $\tau$ measurements to eventually reach the $1\%$ level.  These measurements have the potential to uncover a radiative origin for the $\tau$-lepton and $b$-quark masses. On the other hand, we will demonstrate that even if no deviation is seen, these models could still be part of the solution to the flavour puzzle.

We outline our model building philosophy in \cref{sec:philosophy} and identify the two general classes of models which can generate the $b$-quark and $\tau$-lepton masses at the one-loop level, distinguished by their topologies.  We then classify all models which lead to the first topology for the $\tau$-lepton, the $b$-quark or both in \cref{sec:class1-classification}. We discuss the radiative mass generation in \cref{sec:class1-mass}.  In \cref{sec:class1-pheno} we analyse the phenomenology of three benchmark models in detail, one for the $\tau$-lepton, one for the $b$-quark and one combined model, focusing on the current and future collider constraints on theoretically motivated regions of parameter space.  We then repeat the model classification, radiative mass generation and analysis of benchmark models for the second topology in \cref{sec:class2-classification,sec:class2-mass,sec:class2-pheno}, before concluding. Further details are provided in the appendices.

%=============================================================================
\section{Model building philosophy}
\label{sec:philosophy}
%=============================================================================

As outlined above, our approach is to construct general frameworks for the radiative generation of the $b$-quark and $\tau$-lepton masses and effective Yukawa couplings to the Higgs boson. The observed values for $m_b$ and $m_\tau$ give $m_b/v \sim m_\tau/v \sim 0.01$, indicating that a one-loop suppression ($1/16\pi^2$) is appropriate for generating the observed $b$-quark and $\tau$-lepton masses with order one coupling constants. As such, we consider models which can generate these masses at the one-loop level.  We remain agnostic about how the first- and second-family charged fermions and the neutrinos obtain their masses, and assume the required extended dynamics does not significantly modify our conclusions from the third-family analysis. The observed structure of the CKM matrix and the strong bounds on charged lepton flavour violation suggest that this is a reasonable assumption.
Detailed model-building would be required to produce a phenomenologically-viable and compelling model for all three families and the neutrinos, given that, within the radiative  paradigm, the inter- and intra-family mass hierarchies should have an explanation rather than being due to hierarchies in fundamental parameter values.

\begin{figure}[t]
  \begin{center}
	% !TEX root = ../paper.tex
\noindent\makebox[\columnwidth][c]{
\begin{tabular}{ccc}
\begin{fmffile}{feyngraph-1a}
\begin{fmfgraph*}(120,60)
    \fmfstraight
    \fmfleft{l1,l2}
    \fmfright{r1,r2}
    \fmf{fermion}{l1,v1}
    \fmf{fermion,tension=0.5,label=$\psi$}{v1,v2}
    \fmf{fermion}{v2,r1}
    \fmf{phantom}{l2,v4,r2}
    \fmffreeze
    \fmf{dashes_arrow,right=0.42,label=$\eta^\dagger$}{v3,v1}
    \fmf{dashes_arrow,right=0.42,label=$\phi^\dagger$}{v2,v3}
    \fmf{dashes_arrow,tension=2.05}{v4,v3}
    \fmflabel{$\tau_R$, $b_R$}{l1}
    \fmflabel{$H$}{v4}
    \fmflabel{$L_L$, $Q_L$}{r1}
  \end{fmfgraph*}
\end{fmffile}
& \hspace{3cm} &
\begin{fmffile}{feyngraph-1b}
\begin{fmfgraph*}(120,60)
    \fmfstraight
    \fmfleft{l1,l2}
    \fmfright{r1,r2}
    \fmf{fermion}{l1,v1}
    \fmf{dashes_arrow,tension=0.5,label=$\phi^\dagger$,label.side=left}{v2,v1}
    \fmf{fermion}{v2,r1}
    \fmf{phantom}{l2,v4,r2}
    \fmffreeze
    \fmf{fermion,left=0.42,label=$\chi$}{v1,v3}
    \fmf{fermion,left=0.42,label=$\psi$}{v3,v2}
    \fmf{dashes_arrow,tension=2.05}{v4,v3}
    \fmflabel{$\tau_R$, $b_R$}{l1}
    \fmflabel{$H$}{v4}
    \fmflabel{$L_L$, $Q_L$}{r1}
  \end{fmfgraph*}
\end{fmffile}
\\\\
Class 1 && Class 2
\end{tabular}
}
  \end{center}
	\caption{One-loop diagrams that generate effective $\tau$-lepton and $b$-quark Yukawa couplings after integrating out the heavy, exotic fields.}
	\label{fig:feyn}
\end{figure}
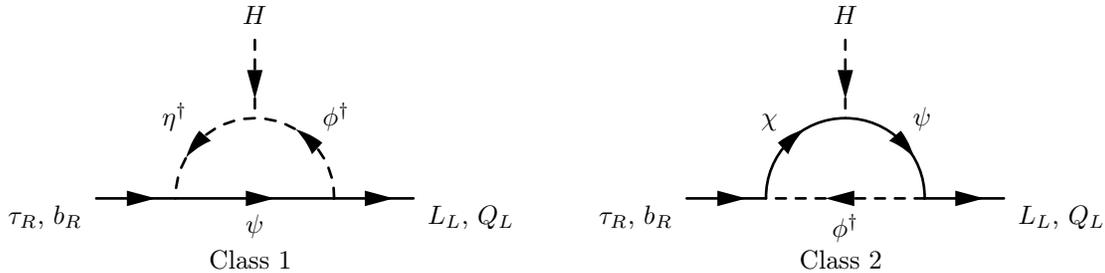

In constructing these frameworks we follow the approach of minimality and introduce only the necessary features required to generate the fermion masses and Yukawa couplings.  We also favour models that introduce the fewest number of new degrees of freedom.  For simplicity, we do not consider extended gauge structures and so restrict the new particles we introduce to be either massive scalars or vector-like (or possibly Majorana) fermions.  Within these constraints, there are two one-loop topologies which can generate effective $b$ and $\tau$ Yukawa couplings, shown in \cref{fig:feyn}, which we term Class~1 and Class~2.  

Class~1 models feature a vector-like exotic fermion $\psi$ and two exotic scalars $\eta$ and $\phi$, while Class~2 models feature two vector-like fermions $\chi$ and $\psi$ and one scalar $\phi$. We assume that none of the exotic scalars acquire a vev, since this would lead to tree-level fermion mass generation (and in specific models may also result in a charge- or colour-breaking vacuum). Note that while we use the labels $\psi$ and $\phi$ for particles in both Class~1 and Class~2 models, these particles are distinct and their quantum numbers are not in general the same.  

Both classes have symmetries that forbid the tree-level SM Yukawa coupling and which are broken only softly. These symmetries are broken by the scalar trilinear coupling and fermion mass term in Class~1 and fermion mass terms in Class~2. Soft breaking via scalar mass terms does not occur in the minimal models we consider, although can do in non-minimal models with additional scalars. 

In this paper we first consider Class~1 models, followed by Class~2.  We do not consider models where both topologies are generated.  For each topology we first classify the models and then perform a detailed phenomenological analysis of several benchmark models:~one which radiatively generates the $\tau$-lepton mass, one for the $b$-quark, and one which can generate both simultaneously. 

%========================================================
%=========================================================

%=============================================================================
\section{Class~1 -- Model Classification}
\label{sec:class1-classification}
%=============================================================================

We first classify Class~1 models which can generate the mass and Yukawa coupling of the $\tau$-lepton, followed by the $b$-quark, and then models which can generate the masses of both the $\tau$-lepton and the $b$-quark at the same time.  These combined models are particularly economical since they have the same number of new degrees of freedom as the analogous models for the $\tau$-lepton or $b$-quark alone.  We then highlight the models which contain only singlets and fundamentals of the SM gauge group, from which we take the benchmark models in \cref{sec:class1-pheno}.

%==============================================================================
\subsection{\texorpdfstring{Class~1 $\tau$-lepton Model Classification}{Class~1 tau-lepton Model Classification}}
%==============================================================================

The Lagrangian exhibiting the relevant terms is
\begin{equation}
    \mathcal{L}_\tau \supset\, - y_\phi \overline{L}_L \phi^\dagger \psi_R - y_\eta \overline{\psi}_L \eta \tau_R - a H \eta^\dagger \phi - m_\psi \overline{\psi}_L \psi_R + \mathrm{h.c.}
        \label{eq:Lag1tau}
\end{equation}
where $L_L$ is the third-family left-handed lepton doublet, $\tau_R$ is the right-handed $\tau$ lepton, $H$ is the Higgs doublet, and $\psi$, $\phi$ and $\eta$ are the exotic fields from the left diagram in \cref{fig:feyn}. We can take all couplings to be real and positive without loss of generality. The tree-level $\tau$ Yukawa interaction is required to be absent. We now analyse the symmetries of this theory, and identify how the tree-level $\tau$ Yukawa term is forbidden.

A simple analysis of the possible $U(1)$ symmetries reveals three independent generators. The first is hypercharge $Y$, with the SM fields having their usual assignments, augmented by
\begin{equation}
    Y_\phi = \frac{1}{2} + Y_\psi,\qquad Y_\eta = 1 + Y_\psi ,
    \label{eq:Yexotics}
\end{equation}
where $Y_f$ is the hypercharge of field $f$ (we normalise such that electric charge $Q = I_3 + Y$). Notice that there is one free parameter, which we can take to be $Y_\psi$. It is interesting to identify special cases where additional Lagrangian terms fix this parameter to definite values. 

The second generator is lepton number $L$. It is convenient to assign the $L$ charges of the exotics such that $\psi$ has $L=0$ while $\phi$ and $\eta$ have $L=-1$.

The third generator is exotic particle number, which we denote by $X$. All SM fields are neutral under $X$, while $\psi$, $\phi$ and $\eta$ have equal values which we normalise to $X=1$. This symmetry has an interesting role to play, because if it remains unbroken then it enforces stability of the lightest exotic particle. If that state is electrically neutral and a colour singlet, then it is a dark matter (DM) candidate. On the other hand, if the lightest exotic is electrically charged or coloured, then it is subject to strong constraints.

The $U(1)_X$ symmetry is anomaly-free, so it may also be gauged (although we do not do so). The reason for the arbitrary parameter $Y_\psi$ in the hypercharge assignments for the exotics is now also clear.  We can define a ``standard hypercharge $Y_\textrm{st}$'' with the usual values for the SM fields together with $1/2$, $1$ and $0$ for $\phi$, $\eta$ and $\psi$, respectively, but the actual hypercharge is permitted to be the linear combination $Y_\textrm{st} + Y_\psi X$, which reproduces \cref{eq:Yexotics}. Since both $Y_\textrm{st}$ and $X$ are anomaly-free, all linear combinations of them are also anomaly-free.

None of these three symmetries forbids the tree-level $\tau$ Yukawa term. However, there are two softly broken symmetries that enforce the absence of this term. These two symmetries are easily identified by examining the two terms in \cref{eq:Lag1tau} that have the capacity to be explicit, but soft, breaking terms:~the trilinear $a$ term and the $\psi$ mass term. The softly-broken symmetries are revealed by omitting each of these terms in turn.

When the $a$-term is absent, there is an additional $U(1)$ that can be taken to be
\begin{equation}
    \tau_R \to e^{i\theta} \tau_R,\quad  \eta \to e^{- i \theta} \eta,
\end{equation}
with all other fields neutral. Since $\tau_R$ transforms under it, while $L_L$ and $H$ do not, the tree-level $\tau$-Yukawa term is forbidden. Being softly-broken, this approximate symmetry ensures that the $\tau$-Yukawa is not generated by any divergent radiative correction, so no counter-term is necessary. We denote the generator of this symmetry by $S_a$ normalised so that $\tau_R$ has $S_a = 1$.

When $m_\psi = 0$, the softly broken $U(1)$ can be identified with
\begin{equation}
    \tau_R \to e^{i\theta} \tau_R,\qquad \psi_L \to e^{i\theta} \psi_L,
\end{equation}
with all other fields neutral. Once again, it clearly forbids the $\tau$-Yukawa term. Call the generator $S_\psi$ with these two fields having $S_\psi = 1$.

%============================================================
\begin{table}[t]
    \begin{subtable}[h]{0.35\textwidth}
        \begin{tabular}{@{\hspace{1em}} c @{\hspace{2em}} c @{\hspace{2em}} c @{\hspace{1em}}}
            \toprule
            & $SU(3)_C$ & $SU(2)_L$ \\ 
            \midrule
            $\psi$ & $(a,b)$ & $(c)$ \\
            $\phi$ & $(a,b)$ & $(|c \pm 1|)$ \\
            $\eta$ & $(a,b)$ & $(c)$ \\
            \bottomrule
        \end{tabular}
        \caption{$\tau$-lepton}
        \label{tab:class1tau-DynkinLabels}
    \end{subtable}
    \begin{subtable}[h]{0.55\textwidth}
        \begin{tabular}{@{\hspace{1em}} c @{\hspace{2em}} c c c     @{\hspace{2em}} c @{\hspace{1em}}}
            \toprule
            & \multicolumn{3}{c}{$SU(3)_C$} \hspace{1em} & $SU(2)_L$ \\
            & I & II & III & \\  
            \midrule
            $\psi$ & $(a,b)$ & $(a+1,b)$ & $(a,b+1)$ & $(c)$ \\
            $\phi$ & $(a,b+1)$ & $(a,b)$ & $(a+1,b)$ & $(|c \pm 1|)$ \\
            $\eta$ & $(a,b+1)$ & $(a,b)$ & $(a+1,b)$ & $(c)$ \\
            \bottomrule
        \end{tabular}
        \caption{$b$-quark}
        \label{tab:class1b-DynkinLabels}
    \end{subtable}
    \caption{Dynkin labels giving the allowed colour and weak-isospin assignments of the new fields in Class~1 for the $\tau$-lepton (left) and $b$-quark (right).  For the $b$-quark models there are three possible patterns for the $SU(3)_C$ assignments.}
    \label{tab:class1-DynkinLabels}
\end{table}
%============================================================

{\it A priori}, there is an infinite number of colour and weak-isospin assignments for the exotic fields, given in terms of Dynkin labels in \cref{tab:class1-DynkinLabels} (left). We are mainly interested in the lowest-dimensional assignments in order to minimise the number of new degrees of freedom. The special cases involving just singlet and fundamental representations are given in \cref{tab:class1-tau}, together with all the Abelian charges discussed above.

%=====================================================================
\begin{table}[t]
    \begin{tabular}{@{\hspace{1em}} c @{\hspace{2em}} c @{\hspace{2em}} c @{\hspace{2em}} c @{\hspace{2em}} c @{\hspace{2em}} c @{\hspace{2em}} c @{\hspace{2em}} c @{\hspace{1em}}}
        \toprule
        & $L_L$ & $\tau_R$ & $H$ & $\psi_L$ & $\psi_R$ & $\phi$ & $\eta$ \\
        \midrule
        $Y$ & $-\frac{1}{2}$ & $-1$ & $\frac{1}{2}$ & $Y_\psi$ & $Y_\psi$ & $Y_\psi\! +\! \frac{1}{2}$ & $Y_\psi\! +\! 1$ \\
        $L$ & $1$ & $1$ & $0$ & $0$ & $0$ & $-1$ & $-1$\\
        $X$ & $0$ & $0$ & $0$ & $1$ & $1$ & $1$ & $1$ \\
        \midrule
        $S_\psi$ & $0$ & $1$ & $0$ & $1$ & $0$ & $0$ & $0$ \\
        $S_a$ & $0$ & $1$ & $0$ & $0$ & $0$ & $0$ & $-1$ \\
        \midrule
        $SU(3)_C\! \times\! SU(2)_L$ & $(1,2)$ & $(1,1)$ & $(1,2)$ & $(1,1)$ & $(1,1)$ & $(1,2)$ & $(1,1)$ \\
        &&&& $(1,2)$ & $(1,2)$ & $(1,1)$ & $(1,2)$ \\
        &&&& $(3,1)$ & $(3,1)$ & $(3,2)$ & $(3,1)$ \\
        &&&& $(3,2)$ & $(3,2)$ & $(3,1)$ & $(3,2)$ \\
        &&&& $(3^*,1)$ & $(3^*,1)$ & $(3^*,2)$ & $(3^*,1)$ \\
        &&&& $(3^*,2)$ & $(3^*,2)$ & $(3^*,1)$ & $(3^*,2)$ \\
        \bottomrule
    \end{tabular}
    \caption{Quantum numbers of fields for Class~1 $\tau$-lepton models.  The last six lines list the models containing only singlets and fundamentals under the SM gauge group.}
    \label{tab:class1-tau}
\end{table}
%============================================================================

%==============================================================================
\subsection{\texorpdfstring{Class~1 $b$-quark Model Classification}{Class~1 b-quark Model Classification}}
%==============================================================================

The results are very similar for these models. The relevant Lagrangian terms are
\begin{equation}
        \mathcal{L}_b \supset\, - y_t \overline{Q}_L \widetilde{H} t_R -  y_\phi \overline{Q}_L \phi^\dagger \psi_R - y_\eta \overline{\psi}_L \eta b_R - a H \eta^\dagger \phi - m_\psi \overline{\psi}_L \psi_R + \mathrm{h.c.}
\label{eq:Lag1b}
\end{equation}
where $Q_L$ is the third-family left-handed quark doublet, $t_R$ and $b_R$ are the right-handed top and bottom quarks, and the exotics are as per the left-hand diagram in \cref{fig:feyn}. This Lagrangian features a tree-level top-quark Yukawa term, but no such term for the bottom quark.

There are again three exact Abelian generators.  The first is hypercharge with
\begin{equation}
    Y_\phi = -\frac{1}{6} + Y_\psi,\quad Y_\eta = \frac{1}{3} + Y_\psi\,.
\end{equation}
The second is baryon number $B$ with the usual assignments for the SM fields, together with $B=0$ for $\psi$, and with $\phi$ and $\eta$ having $B=-1/3$. The third generator is, again, the anomaly-free exotic particle number $X$. The non-Abelian multiplet structures are given in \cref{tab:class1-DynkinLabels} (right). 

As with the $\tau$-lepton case, these three exact symmetries do not forbid the tree-level $b$ Yukawa term. Its omission is instead justified by two softly-broken $U(1)$ symmetries, identified by omitting the two soft terms proportional to $a$ and $m_\psi$ in turn. Omitting the $a$-term reveals $U(1)_{S_a}$ with
\begin{equation}
     b_R \to e^{i\theta} b_R,\quad \eta \to e^{ - i \theta} \eta,
\end{equation}
while putting $m_\psi = 0$ reveals the symmetry $U(1)_{S_\psi}$
\begin{equation}
    b_R \to e^{i\theta} b_R,\quad \psi_L \to e^{i\theta} \psi_L,
\end{equation}
while all other fields do not transform. 

The minimal models are summarised in \cref{tab:class1-bquark}.

%=============================================================
\begin{table}[t]
    \begin{tabular}{@{\hspace{1em}} c @{\hspace{2em}} c @{\hspace{2em}} c @{\hspace{2em}} c @{\hspace{2em}} c @{\hspace{2em}} c @{\hspace{2em}} c @{\hspace{2em}} c @{\hspace{2em}} c @{\hspace{1em}}}
        \toprule
        & $Q_L$ & $t_R$ & $b_R$ & $H$ & $\psi_L$ & $\psi_R$ & $\phi$ & $\eta$ \\
        \midrule
        $Y$ & $\frac{1}{6}$ & $\frac{2}{3}$ & -$\frac{1}{3}$ & $\frac{1}{2}$ & $Y_\psi$ & $Y_\psi$ & $Y_\psi\! -\! \frac{1}{6}$ & $Y_\psi\! +\! \frac{1}{3}$ \\
        $B$ & $\frac{1}{3}$ & $\frac{1}{3}$ & $\frac{1}{3}$ & $0$ & $0$ & $0$ & $-\frac{1}{3}$ & $-\frac{1}{3}$\\
        $X$ & $0$ & $0$ & $0$ & $0$ & $1$ & $1$ & $1$ & $1$ \\
        \midrule
        $S_\psi$ & $0$ & $0$ & $1$ & $0$ & $1$ & $0$ & $0$ & $0$ \\
        $S_a$ & $0$ & $0$ & $1$ & $0$ & $0$ & $0$ & $0$ & $-1$ \\
        \midrule
        $SU(3)_C\! \times\! SU(2)_L$ & $(3,2)$ & $(3,1)$ & $(3,1)$ & $(1,2)$ & $(1,1)$ & $(1,1)$ & $(3^*,2)$ & $(3^*,1)$ \\
        &&&&& $(1,2)$ & $(1,2)$ & $(3^*,1)$ & $(3^*,2)$ \\
        &&&&& $(3,1)$ & $(3,1)$ & $(1,2)$ & $(1,1)$ \\
        &&&&& $(3,2)$ & $(3,2)$ & $(1,1)$ & $(1,2)$ \\
        &&&&& $(3^*,1)$ & $(3^*,1)$ & $(3,2)$ & $(3,1)$ \\
        &&&&& $(3^*,2)$ & $(3^*,2)$ & $(3,1)$ & $(3,2)$ \\
        \bottomrule
    \end{tabular}
    \caption{Quantum numbers of fields for Class~1 $b$-quark models. The last six lines list the models containing only singlets and fundamentals under the SM gauge group.}
    \label{tab:class1-bquark}
\end{table}
%==============================================================

%==============================================================================
\subsection{\texorpdfstring{Class~1 Combined $b+\tau$ Model Classification}{Class~1 Combined b + tau Model Classification}}
\label{sec:class1-classification-combined}
%==============================================================================

We now consider combined models. To define the possible field identifications it is convenient to rename the exotic fields in \cref{tab:class1-tau} as $\psi_\tau$, $\phi_\tau$ and $\eta_\tau$, and similarly add the subscript $b$ to the exotics in \cref{tab:class1-bquark}. 

It is clear from \cref{tab:class1-tau,tab:class1-bquark} that there are many opportunities to identify either the exotic fermions or the exotic scalars as the same particles in both effective Yukawa diagrams. \emph{A priori}, there are six general cases:
\begin{enumerate}
    \item $\psi_\tau = (\psi_b)^c$.
    \item $\psi_\tau = \psi_b$.
    \item $\phi_\tau = \phi_b$ and $\eta_\tau = \eta_b$.
    \item $\phi_\tau = \eta_b^*$ and $\eta_\tau = \phi_b^*$.
    \item $\phi_\tau = \eta_b$ and $\eta_\tau = \phi_b$.
    \item $\phi_\tau = \phi_b^*$ and $\eta_\tau = \eta_b^*$.
\end{enumerate}
However, cases 5 and 6 can be immediately seen to be less interesting because the hypercharge assignments required for both field identifications within each case are incompatible. Thus only one of the two field identifications in each of these two cases is possible, and so no great gain in economy is obtained.  It is straightforward to check that cases $1$-$4$ do not have any inconsistency issues. It turns out that cases 1 and 3 have a very interesting overlap, leading to the most economical models. We therefore discuss these scenarios in more detail.

We start with the identification $\psi_\tau = (\psi_b)^c$. Given the $SU(3)_C$ representations in \cref{tab:class1-DynkinLabels}, this identification generally requires the associated scalars to be distinct. However, the subset of models where $\psi_b \sim (a,a+1)$ and $\phi_b, \eta_b \sim (a+1,a)$, pattern III in \cref{tab:class1-DynkinLabels} (right) with $a=b$, also allow for the scalar identifications $\phi_\tau=\phi_b$ and $\eta_\tau=\eta_b$. The consistent hypercharge assignment is $Y_{\psi_b} = -Y_{\psi_\tau} = 1/3$.

The Lagrangians for these models have the form
\begin{align}
\label{eq:Lag1comb}
    \mathcal{L}_{b\tau} \subset\, &- y_t \overline{Q}_L \widetilde{H} t_R -  y_\phi^b\, \overline{Q}_L \phi^\dagger \psi_R - y_\phi^\tau\, \overline{L}_L \phi^\dagger (\psi_L)^c - y_\eta^b\, \overline{\psi}_L\eta\, b_R - y_\eta^\tau\, \overline{(\psi_R)^c}\, \eta\, \tau_R \nonumber\\
    &- a H \eta^\dagger \phi - m_\psi \overline{\psi}_L\, \psi_R + \mathrm{h.c.}
\end{align}
There are two conserved charges in addition to hypercharge, which are given by
\begin{equation}
    B + \tfrac{1}{6} X'\quad \textrm{and}\quad L - \tfrac{1}{2} X ,
\end{equation}
where
\begin{equation}
    X_\psi = X_\phi = X_\eta = 1\quad \textrm{and}\quad X'_\psi = - X'_\phi = - X'_\eta = 1 .
\end{equation}
These are clearly extensions of the usual baryon and lepton numbers to take account of interactions with the exotic sector.  In addition to the above symmetries, there is a $\mathcal{Z}_2$ parity under which the exotic fields are odd. 

The softly broken symmetries that ensure the absence of the tree-level $b$ and $\tau$ Yukawa interactions are defined by the following charge assignments under $S_a$ and $S_\psi$:
\begin{equation}
    S_a(b_R) = S_a(\tau_R) = - S_a(\eta) = 1 \,,
\end{equation}
and
\begin{equation}
    S_\psi(b_R) = S_\psi(\tau_R) = 2,\quad S_\psi(\psi_L) = - S_\psi(\psi_R) = -S_\psi(\phi) = - S_\psi(\eta) = 1 \,,
\end{equation}
with all other fields neutral. 

The minimal combined models are summarised in \cref{tab:class1-combined}. Notice that in these models $\psi^c$ has the gauge quantum numbers of a down-type vector-like quark; however, mixing with the SM quarks is forbidden by the $B$ and $L$ symmetries.

%=====================================================================
\begin{table}[ht]
    \begin{tabular}{@{\hspace{0.6em}} c @{\hspace{1.2em}} c @{\hspace{1.5em}} c @{\hspace{1.2em}} c @{\hspace{1.2em}} c @{\hspace{1.2em}} c @{\hspace{1.2em}} c @{\hspace{1.2em}} c @{\hspace{1.2em}} c @{\hspace{1.2em}} c @{\hspace{1.2em}} c @{\hspace{1.2em}} c @{\hspace{0.6em}}}
        \toprule
        & $Q_L$ & $b_R$ & $L_L$ & $\tau_R$ & $H$ & $\psi_L$ & $\psi_R$ & $\phi$ & $\eta$ \\
        \midrule
        $Y$ & $\frac{1}{6}$ & $-\frac{1}{3}$ & $-\frac{1}{2}$ & $-1$ & $\frac{1}{2}$ & $\frac{1}{3}$ & $\frac{1}{3}$ & $\frac{1}{6}$ & $\frac{2}{3}$ \\
        $B$ & $\frac{1}{3}$ & $\frac{1}{3}$ & $0$ & $0$ & $0$ & $\frac{1}{6}$ & $\frac{1}{6}$ & $-\frac{1}{6}$ & $-\frac{1}{6}$ \\
        $L$ & $0$ & $0$ & $1$ & $1$ & $0$ & $-\frac{1}{2}$ & $-\frac{1}{2}$ & $-\frac{1}{2}$ & $-\frac{1}{2}$ \\
        $\mathcal{Z}_2$ & 1 & 1 & 1 & 1 & 1 & $-1$ & $-1$ & $-1$ & $-1$ \\
        \midrule
        $S_\psi$ & $0$ & $2$ & $0$ & $2$ & $0$ & $1$ & $-1$ & $-1$ & $-1$ \\
        $S_a$ & $0$ & $1$ & $0$ & $1$ & $0$ & $0$ & $0$ & $0$ & $-1$ \\
        \midrule
        $SU(3)_C\! \times\! SU(2)_L$ & $(3,2)$ & $(3,1)$ & $(1,2)$ & $(1,1)$ & $(1,2)$ & $(3^*,1)$ & $(3^*,1)$ & $(3,2)$ & $(3,1)$ \\
        &&&&&& $(3^*,2)$ & $(3^*,2)$ & $(3,1)$ & $(3,2)$ \\
        \bottomrule
    \end{tabular}
    \caption{Quantum numbers of the fields for the minimal Class~1 combined models. The last two lines list the models containing only singlets and fundamentals under the SM gauge group.}
    \label{tab:class1-combined}
\end{table}
%============================================================================

%==============================================================================
\section{Class~1 -- Radiative Mass Generation}
\label{sec:class1-mass}
%==============================================================================

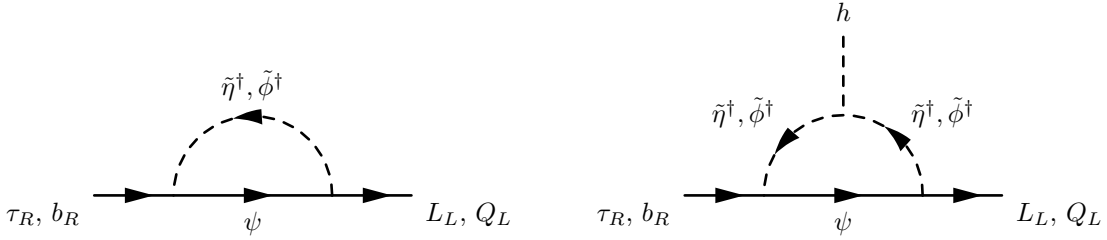
\begin{figure}[t]
  \begin{center}
	% !TEX root = ../paper.tex
\noindent\makebox[\columnwidth][c]{
\begin{tabular}{ccc}
\begin{fmffile}{feyngraph-3a}
\begin{fmfgraph*}(120,60)
    \fmfstraight
    \fmfleft{l1,l2}
    \fmfright{r1,r2}
    \fmf{fermion}{l1,v1}
    \fmf{fermion,tension=0.5,label=$\psi$}{v1,v2}
    \fmf{fermion}{v2,r1}
    \fmf{phantom}{l2,v4,r2}
    \fmffreeze
    \fmf{dashes_arrow,right,label=$\tilde{\eta}^\dagger,,\tilde{\phi}^\dagger$}{v2,v1}
    \fmflabel{$\tau_R$, $b_R$}{l1}
    \fmflabel{$L_L$, $Q_L$}{r1}
  \end{fmfgraph*}
\end{fmffile}
& \hspace{3cm} &
\begin{fmffile}{feyngraph-3b}
\begin{fmfgraph*}(120,60)
    \fmfstraight
    \fmfleft{l1,l2}
    \fmfright{r1,r2}
    \fmf{fermion}{l1,v1}
    \fmf{fermion,tension=0.5,label=$\psi$}{v1,v2}
    \fmf{fermion}{v2,r1}
    \fmf{phantom}{l2,v4,r2}
    \fmffreeze
    \fmf{dashes_arrow,right=0.42,label=$\tilde{\eta}^\dagger,,\tilde{\phi}^\dagger$}{v3,v1}
    \fmf{dashes_arrow,right=0.42,label=$\tilde{\eta}^\dagger,,\tilde{\phi}^\dagger$}{v2,v3}
    \fmf{dashes,tension=2.05}{v4,v3}
    \fmflabel{$\tau_R$, $b_R$}{l1}
    \fmflabel{$h$}{v4}
    \fmflabel{$L_L$, $Q_L$}{r1}
  \end{fmfgraph*}
\end{fmffile}
\end{tabular}
}
  \end{center}
	\caption{One-loop diagrams that generate effective $\tau$-lepton and $b$-quark masses and effective Yukawa couplings after electroweak symmetry breaking in the Class~1 models.}
	\label{fig:feyn-class1-after-ewsb}
\end{figure}

The left diagram of \cref{fig:feyn} generates an effective Yukawa interaction between the Higgs doublet and the $\tau$-lepton (and/or $b$-quark).  Electroweak symmetry breaking mixes the exotic scalar states to produce mass eigenstates $\tilde{\phi}$ and $\tilde{\eta}$.  The $\tau$-lepton (and/or $b$-quark) mass and effective Yukawa coupling to the physical Higgs boson are then generated by the diagrams in \cref{fig:feyn-class1-after-ewsb}.  The relation in \cref{eq:lambda-m-treelevel} between the fermion's mass and its coupling to the physical Higgs is then broken at $\mathcal{O}(v^2/M^2)$, where $M$ is the scale of new physics. This can be understood from the fact that there are additional diagrams which, after integrating out the exotic states, generate operators of the form $\bar{L}_LH\tau_R(H^\dagger H)^n$. 

%========================================================
%=========================================================

We proceed by diagonalising the scalar mass matrices and evaluating the diagrams in \cref{fig:feyn-class1-after-ewsb}.  The mixing between $\phi$ and $\eta$ arises from the $a$-terms in \cref{eq:Lag1b,eq:Lag1tau,eq:Lag1comb}. Since the Higgs field transforms as a doublet of $SU(2)_L$, the dimensions of the representations of $\phi$ and $\eta$ differ by one (see \cref{tab:class1-DynkinLabels}).  This means that the larger $SU(2)_L$ multiplet contains one state that does not mix.  We write the representation of the smaller multiplet as $n_2\equiv\min(d_2(\phi),d_2(\eta))$, where $d_2$ 
is the dimension of the $SU(2)_L$ representation.  The states that mix all do so with the same mixing angle $\theta$.  The mass eigenstates are then
\begin{equation}
    \begin{pmatrix}
        \tilde{\phi} \\
        \tilde{\eta}
    \end{pmatrix}
    =
    \begin{pmatrix}
        \cos\theta & -\sin\theta \\
        \sin\theta & \cos\theta
    \end{pmatrix}
    \begin{pmatrix}
        \phi \\
        \eta
    \end{pmatrix}
    \,,
\end{equation}
with $\theta \in [0,\tfrac{\pi}{2}]$. The corresponding eigenvalues are
\begin{align}
    m_1^2 \equiv m_{\tilde{\phi}}^2 &=\, \frac{1}{2}\left(m_\eta^2 + m_\phi^2 - \sqrt{(m_\eta^2 - m_\phi^2)^2+2 a^2 v^2}\right) \,, \\
    m_2^2 \equiv m_{\tilde{\eta}}^2 &=\, \frac{1}{2}\left(m_\eta^2 + m_\phi^2 + \sqrt{(m_\eta^2 - m_\phi^2)^2+2 a^2 v^2}\right) \,,
\end{align}
where we have defined $m_1=m_{\tilde{\phi}}$ and $m_2=m_{\tilde{\eta}}$. Note that $m_1 \leq m_2$. The mixing angle is given by
\begin{equation} \label{eq:mixing1}
    \sin(2\theta) = \frac{\sqrt{2} a v}{m_2^2 - m_1^2} \,,
\end{equation}
with $\theta \in [0,\tfrac{\pi}{4}]$ for $m_\phi \leq m_\eta$ and $\theta \in [\tfrac{\pi}{4},\tfrac{\pi}{2}]$ for $m_\phi \geq m_\eta$. This implies that $m_2^2 -m_1^2 \geq \sqrt{2}av$, with equality at maximal mixing ($\theta = \tfrac{\pi}{4}$).

We can now calculate the one-loop $b$-quark and $\tau$-lepton masses, \cref{fig:feyn-class1-after-ewsb} (left), which are given by
\begin{align} \label{eq:mf1}
    m_{\tau,b} = \frac{y_\phi y_\eta}{16\pi^2} 
    \frac{v}{\sqrt{2}}
    \frac{a m_\psi}{m_1 m_2} n_2 n_3 F\left(x_1, x_2\right) \,,
\end{align}
where $x_{1,2} = m_{1,2}^2/m_\psi^2$, $n_2$ is again the dimension of the smaller $SU(2)_L$ multiplet of $\phi$ and $\eta$, and $n_3$ is a model-dependent $SU(3)_C$ group theory factor. Note that \cref{eq:mixing1} has been used to remove the dependence on the mixing angle. The loop function is 
\begin{equation} \label{eq:F}
    F(x_1,x_2) = \frac{\sqrt{x_1 x_2}}{x_1-x_2} \left( \frac{x_1}{x_1-1}\ln x_1 - \frac{x_2}{x_2-1}\ln x_2 \right) \,,
\end{equation}
and satisfies $0 \leq F(x_1,x_2) \leq 1$.

Notice that the generated mass, \cref{eq:mf1}, is invariant under simultaneous rescalings of $a$, $m_\psi$, $m_1$ and $m_2$; in other words, the fermion masses do not depend on the overall scale of new physics. This lack of decoupling arises from the fact that the SM Yukawa couplings are dimensionless or, equivalently, that we are computing contributions to the Wilson coefficients of marginal operators. At first glance this seems problematic, since the new physics could be arbitrarily heavy; however, naturalness arguments provide a strong motivation for lower mass scales. Specifically, there are one-loop contributions to the Higgs mass that lead to fine-tuning of the electroweak scale if $a$ is too large. This is discussed in further detail in \cref{app:naturalness}.

Next, we determine the effective Yukawa couplings between the fermions and the physical Higgs boson, \cref{fig:feyn-class1-after-ewsb} (right).  These effective Yukawa couplings are momentum-dependent, and for the decay of an on-shell Higgs boson should be evaluated at $p_h^2 = m_h^2$. While we do this in our numerical results, the functional forms are unwieldy and so here we instead present the expression at $p_h^2=0$ (the difference is in any case negligible for new physics significantly heavier than $m_h$).  The expression then takes the simpler form\footnote{Note that while our radiative mass, \cref{eq:mf1}, agrees with the result found for the model studied in~\cite{Fraser:2014ija}, we find a different form for the effective Yukawa coupling.}
\begin{align} \label{eq:yeff1}
    y_{\tau,b}^\text{eff}(p_h^2=0) &= \frac{y_\phi y_\eta}{16 \pi^2} a m_\psi n_2 n_3 \left[ \cos^2(2\theta) \frac{F(x_1,x_2)} {m_1 m_2} + \frac{1}{2} \sin^2(2\theta) \left( \frac{F(x_1)}{m_1^2} + \frac{F(x_2)}{m_2^2} \right) \right] \,,
\end{align}
where again $x_{1,2} = m_{1,2}^2/m_\psi^2$ and we have defined
\begin{align}
    F(x) \equiv \lim_{y\to x}F(x,y)
    = \frac{x}{x-1}-\frac{x\log(x)}{(x-1)^2}\,.
\end{align}
As noted above, radiative models of fermion mass generation do not satisfy the SM relation $m_f=y_f v/\sqrt{2}$, as can be seen from \cref{eq:mf1,eq:yeff1}. This relation is, however, approximately satisfied if either $\sin(2\theta)$ is small (i.e., if $|m_\eta^2-m_\phi^2| \gg \sqrt{2}av$) or if $\sin(2\theta) \approx 1$ and $m_1 \approx m_2$ (i.e., if $|m_\eta^2-m_\phi^2| \ll 2a^2v^2$ and $m_\eta^2 \approx m_\phi^2 \gg av$). Furthermore, we have the prediction that $y_f \geq \sqrt{2} m_f/v$ in these models.

%==============================================================================
\section{Class~1 -- Benchmark models}
\label{sec:class1-pheno}
%==============================================================================

We now turn to exploring the phenomenology of this class of models. For illustrative purposes, we consider three benchmark models; however, many of the features are applicable more broadly.

As highlighted in the introduction, the most important prediction of radiative mass generation models is that they violate the SM tree-level relation between a fermion's mass and its coupling to the physical Higgs boson. Precision Higgs measurements therefore provide an effective way to test these models. In addition to an increase in the $\tau$-lepton/$b$-quark couplings, there are new one-loop contributions to the Higgs couplings to gauge bosons. These are most relevant for the gluon and photon couplings, where there is no tree-level coupling in the SM. These new contributions interfere constructively with the SM in the case of the photon and destructively for the gluon. The modification of the effective Higgs-gluon vertex can then either enhance or suppress the gluon-fusion production cross-section, depending on the relative sizes of the new physics and SM contributions, potentially impacting a wide range of Higgs measurements.  We calculate these contributions and present the current signal strength measurements in \cref{app:higgs-couplings}.  

These models also affect a number of observables that enter the electroweak fit, since there are new states charged under $SU(2)_L \times U(1)_Y$. These effects can be largely taken into account via the $S$ and $T$ parameters, provided that the new physics is above the weak scale (which is likely the case due to bounds from direct searches). We calculate the corrections for all Class~1 models in \cref{app:STU} and present the current experimental limits.  There are also non-oblique corrections in the form of modified couplings of the $Z$-boson to the $\tau$-lepton/$b$-quark, discussed in \cref{app:Z-couplings} . We find that these corrections are generally smaller than the current experimental uncertainties, but could be observable in a $Z$-pole run at a future $e^+e^-$ collider.

There are also theoretical constraints on the parameter space of these models.  Firstly, perturbative unitarity requires $y_\phi, y_\eta \lesssim \sqrt{4\pi}$; a significantly stronger bound could be imposed by requiring that the running couplings do not encounter a Landau pole below the Planck scale (or some other intermediate scale).  The trilinear coupling $a$ is also bounded by perturbative unitarity.  We use the \texttt{SARAH}/\texttt{SPheno} framework~\cite{Porod:2003um,Porod:2011nf,Staub:2008uz,Staub:2009bi,Staub:2010jh,Staub:2012pb,Staub:2013tta} to calculate this bound, using the default partial diagonalisation treatment of the $t$- and $u$-channel poles~\cite{Goodsell:2018tti,Goodsell:2020rfu}.  Finally, if $\eta$ or $\phi$ are significantly above the weak scale they generate a fine-tuning problem for the Higgs mass parameter.  This is calculated in \cref{app:naturalness} where we also define our fine-tuning measure.

The effects discussed above are common to all models, although the quantitative predictions depend on the gauge quantum numbers of the states in any given model.  In the subsequent sections we quantitatively analyse these effects, which are directly related to radiative mass generation, but there can be additional interactions in specific models that may provide further phenomenological constraints. Here, and for each benchmark model, we briefly discuss some of this more model-dependent phenomenology.

An important consideration is the presence of the $U(1)_X$ exotic particle number symmetry. If it is unbroken, the lightest state charged under this symmetry will be stable and, if electrically neutral, could provide a dark matter candidate. Whether such a state exists and can produce the required relic abundance depends on the details of each model. On the other hand, if the lightest exotic state carries electric charge or colour there are extremely strong bounds on its cosmological abundance~\cite{0807.0211,1410.1374}. There are two obvious ways to evade such bounds:~(i) extend the exotic sector so the lightest state is neutral, or (ii) break the stabilising symmetry. In certain models the latter can be achieved without introducing any new states, via additional interactions with SM particles.

Finally, there is the possibility of directly producing the new states at colliders. This is especially relevant for the $b$-quark models which necessarily contain new coloured states that can have large production cross-sections at the LHC. The $U(1)_X$ symmetry again plays an important role here. If the stabilising symmetry is preserved, the relevant searches are those with large missing energy. If it is broken, the exotic states could decay promptly into SM particles or be (meta-)stable on collider timescales. In either scenario, the precise collider signatures, and hence the strength of the bounds, are highly model-dependent.

With these general features in mind, we now turn to three benchmark models:~one which generates the $\tau$-lepton mass radiatively, one for the $b$-quark, and a combined model which can radiatively generate both the $\tau$-lepton and the $b$-quark masses simultaneously.

%--------------------------------------------------------
\subsection{\texorpdfstring{Class~1 $\tau$-lepton Benchmark Model}{Class~1 tau-lepton Benchmark Model}}
\label{sec:class1-tau}
%--------------------------------------------------------

For radiative $\tau$-lepton mass and Yukawa generation, we adopt the minimal model (i.e., the model with the smallest gauge representations) as our benchmark model. Namely, 
\begin{equation}
    \psi_{L,R} \sim (\mathbf{1} ,\mathbf{1}, 0)\,, \quad \phi \sim (\mathbf{1}, \mathbf{2}, \tfrac{1}{2}) \,, \quad \eta \sim (\mathbf{1}, \mathbf{1}, 1) \,,
\end{equation}
where we have taken $Y_\psi=0$. For this particular hypercharge assignment, $\psi$ could be a Majorana fermion, minimising the new degrees of freedom introduced. A Majorana mass term would explicitly break $U(1)_X$ to the discrete $\mathcal{Z}_2$ subgroup under which $\psi$, $\phi$ and $\eta$ are odd. Whether $\psi$ is Majorana or Dirac does not significantly impact the electroweak and precision Higgs phenomenology, but for concreteness we continue to assume that $\psi$ is Dirac.  In either case, if $\psi$ is the lightest exotic particle then it can be a dark matter candidate.  Alternatively, this model also allows for additional interactions between the exotics and the SM that break the stabilising symmetry. Examples include $\bar{\psi}_R H L$, $H^\dagger \phi$, or SM Yukawa interactions with $H$ replaced by $\phi$. Note that it is important to ensure that the symmetries $S_\psi$ and $S_a$ are not both explicitly broken, so that the SM $\tau$ Yukawa term remains forbidden.  Finally, the group theory factors in \cref{eq:mf1,eq:yeff1} for this model are $n_2 = 1$ and $n_3 = 1$.

\begin{figure}[ht]
    \centering
    \includegraphics[height=0.45\textwidth]{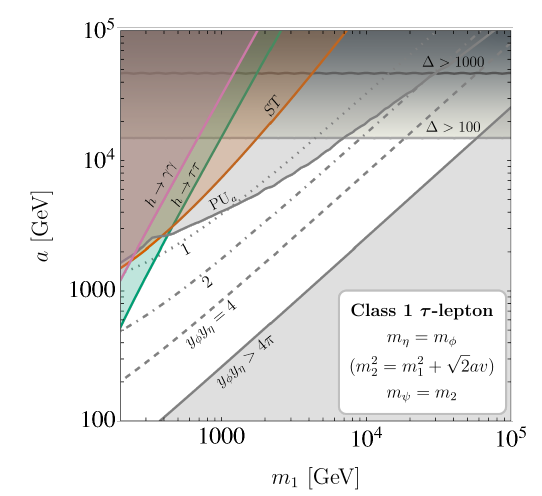}
    \includegraphics[height=0.45\textwidth]{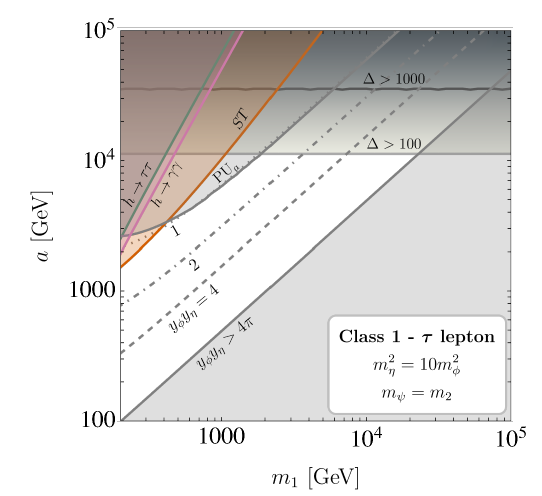}
	\caption{Theoretical (grey) and current $2\sigma$ experimental constraints (coloured) on the Class~1 $\tau$-lepton benchmark model.  The product of Yukawas required to reproduce the observed tau mass is shown by contours at $y_\phi y_\eta \in \{1, 2, 4\}$ and the region which violates perturbative unitarity is shown in grey.  The region where perturbative unitarity of the trilinear coupling is violated is shown in grey, and is indicated by $\text{PU}_a$.  The degree of fine-tuning is shown by contours at $\Delta \in \{100,1000\}$ using the measure defined in \cref{app:naturalness}.  The coloured regions are experimentally excluded at $2\sigma$.  $ST$ represents the exclusion due to the $S$ and $T$ parameters, while $h\to \tau\tau$ and $h\to\gamma\gamma$ give the exclusion due to the indicated signal strength.}
	\label{fig:1tau}
\end{figure}

The constraints on this model are shown in \cref{fig:1tau} for two slices of parameter space.  In the left panel $m_2$ is as light as possible, corresponding to maximal mixing of the scalars, while the right panel fixes $m_\eta^2=10m_\phi^2$.  In both slices the mass of the exotic fermion $\psi$ is taken to be equal to $m_2$.

The theoretical constraints on the model are shown in grey.  The product of Yukawa couplings $y_\phi y_\eta$ which reproduces the measured tau mass is shown by the contours at $y_\phi y_\eta$ = 1, 2, 4 and $4\pi$.  We find that regions of parameter space where $m_1\gg a$ require non-perturbative Yukawas.  Conversely, perturbative unitarity of the trilinear coupling constrains the region where $a\gg m_1$.  Between these constraints, the tau mass can be radiatively generated with order-one Yukawas.  This reflects and validates the observation in \cref{sec:philosophy} that one-loop mass generation is appropriate for the tau since  $m_\tau / v \sim 1/(16\pi^2)$.  Note that the choice $m_\psi=m_2$ approximately maximises the expression for the tau mass in \cref{eq:mf1}, although the dependence on $m_\psi$ is quite mild. Other choices would require larger values of $y_\phi y_\eta$ and hence lead to slightly stronger constraints on the parameter space from perturbativity.  We see that the trilinear coupling and the exotic mass scales can be raised together without running in to perturbativity constraints.  This demonstrates the non-decoupling noted in \cref{sec:class1-mass}.  However, as the scales rise above $10^4\,\text{GeV}$ significant fine-tuning of the electroweak scale becomes necessary, reflected in the large values of $\Delta$ (see \cref{app:naturalness} for our definition of the fine-tuning measure $\Delta$).

The current $2\sigma$ experimental bounds on this model from Higgs and electroweak measurements are shown as coloured regions.  In this model only the $h\bar{\tau}\tau$ and $h\gamma\gamma$ couplings deviate from their SM values.  In the left panel we find that only the Higgs to $\tau^+\tau^-$ signal strength measurement significantly constrains the theoretically allowed parameter space, while in the right panel only the $S$ and $T$ parameters are currently sensitive. Due to the larger value of $m_2$, the experimental bounds are weaker in the right panel, as is the perturbativity bound on $a$ (which is approximately proportional to $m_2$). However, the perturbativity constraint on $y_\phi y_\eta$ and the fine-tuning bound are slightly stronger.  In both slices there remain allowed regions of parameter space with $y_\phi y_\eta = 1$. 

%--------------------------------------------------------
%--------------------------------------------------------

\begin{figure}[ht]
    \centering
    \includegraphics[height=0.45\textwidth]{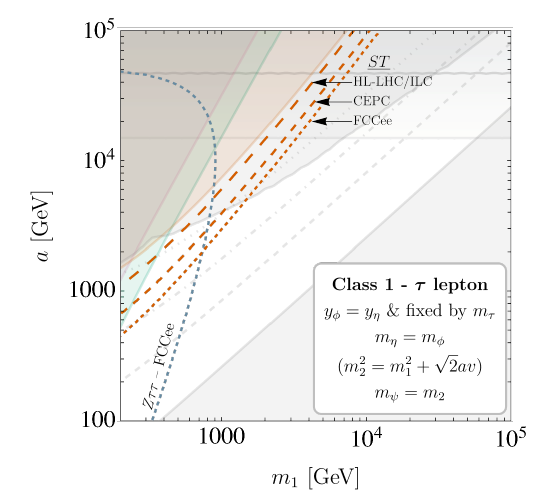}
    \includegraphics[height=0.45\textwidth]{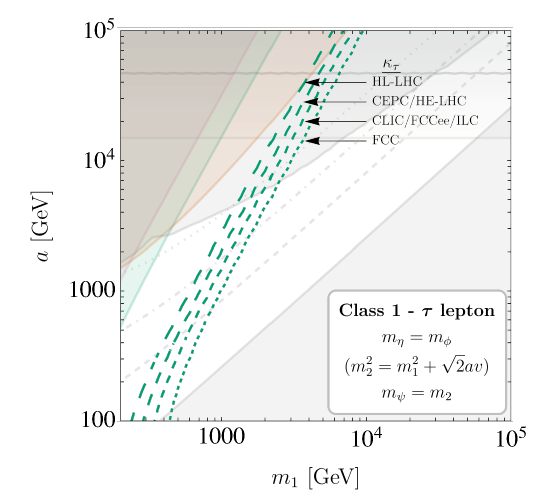}
	\caption{Projected $2\sigma$ experimental sensitivities on the Class~1 $\tau$-lepton benchmark model at future colliders for the left hand slice in \cref{fig:1tau}.  For the $Z\tau\tau$ sensitivity we have assumed $y_\chi = y_\psi$ and fixed the values to reproduce the tau mass.  See \cref{app:higgs-couplings,app:experimental-results} for details of the kappa framework and our assumptions for the running scenarios of future collider experiments.  Where future colliders have comparable sensitivity, we show a single curve for readability.}
	\label{fig:1tau-FC}
\end{figure}

As discussed in \cref{app:experimental-results}, future colliders are expected to significantly increase the precision of both Higgs and electroweak measurements.  The left panel of \cref{fig:1tau-FC} shows that improved sensitivity to $S$ and $T$ can rule out $y_\phi y_\eta = 1$, although slightly larger values would still be allowed.  Improved measurements of the $Z\bar{\tau}\tau$ couplings at the FCC-ee would be able to probe a complementary region of parameter space.  These $Z$ couplings depend on $y_\phi$ and $y_\eta$ independently, rather than the product, and we have fixed $y_\phi = y _\eta$ in the plot.  Other ratios of these Yukawas typically increase the sensitivity.  In the right panel of \cref{fig:1tau-FC}, we present the reach of future determinations of $\kappa_\tau$.  The kappa framework~\cite{LHCHiggsCrossSectionWorkingGroup:2012nn,Heinemeyer:2013tqa} parameterises deviations in Higgs couplings; for more details see \cref{app:higgs-couplings,app:experimental-results}.  Future measurements of $\kappa_\tau$ can provide better sensitivity than $S$ and $T$ for all future experiments we consider.  While future determinations of $\kappa_\gamma$ will probe the parameter space, these are less sensitive than $\kappa_\tau$ and we do not show them.   

%--------------------------------------------------------
%--------------------------------------------------------

In this model the fermion $\psi$ may be a dark matter candidate. The relic abundance is set by annihilation into $\tau^+\tau^-$ and $\nu\bar{\nu}$ via t-channel $\phi$ and $\eta$ exchange. Neglecting the scalar mixing and the SM fermion masses, the thermally-averaged cross-section for Dirac $\psi$ annihilating to taus is
\begin{equation}
    \langle \sigma v \rangle = \frac{1}{32\pi m_\psi^2} \left( \frac{y_\phi^4}{(1 + m_\phi^2/m_\psi^2)^2} + \frac{y_\eta^4}{(1 + m_\eta^2/m_\psi^2)^2} \right) + \mathcal{O}(v^2) \,.
\end{equation}
Given that $\mathcal{O}(1)$ values of $y_\phi$ and $y_\eta$ are needed to radiatively generate the $\tau$ mass, this cross-section can easily be quite large. The observed dark matter abundance could then be obtained with $\psi$ masses up to a few TeV, depending on $m_\phi$ and $m_\eta$.  Direct detection bounds are relatively strong for Dirac dark matter, due to a loop-induced electromagnetic dipole interaction, but disappear for Majorana dark matter, as the dipole interaction then identically vanishes~\cite{Baker:2018uox}.  A detailed study would be required to determine if these scenarios are phenomenologically viable.  While the neutral component of $\phi$ also appears to be a potential dark matter candidate, the mixing with $\eta$ means that one of the charged eigenstates is expected to be lighter.

Since there are no coloured exotics in this model, we do not expect significant constraints from direct production at colliders. Assuming an unbroken $U(1)_X$, $\phi$ and $\eta$ closely resemble left- and right-handed staus, respectively, from a collider viewpoint, allowing us to re-purpose the results from SUSY searches. Assuming $m_\psi\ll m_1\approx m_2$, the current bound is\footnote{There is also an allowed mass region in the range 90-120\,GeV between the LEP and LHC bounds.} $m_1>390$\,GeV~\cite{1911.06660}; this bound becomes weaker with increasing $m_\psi$ and there is no constraint when $m_\psi \gtrsim 150$\,GeV.  Finally, if any of the $U(1)_X$ breaking interactions discussed above are present, then the exotic states will decay into SM particles. Regions of parameter space that would otherwise be excluded by bounds on charged relics or overclosure of the universe then become viable. The presence of these couplings and their relative magnitudes can also significantly impact the collider signatures.

%--------------------------------------------------------
\subsection{\texorpdfstring{Class~1 $b$-quark Benchmark Model}{Class~1 b-quark Benchmark Model}}
%--------------------------------------------------------

In the minimal model for the $\tau$-lepton we found that perturbativity imposed significant restrictions on the parameter space. These constraints will be slightly stronger in the minimal $b$-quark model, due to the fact that $m_b > m_\tau$. For this reason, and for a point of difference, we choose as our benchmark a non-minimal model in which all the exotic states are charged under $SU(3)_C$; the $b$-quark mass is then enhanced by a colour factor. Specifically, we take
\begin{equation}
    \psi_{L,R} \sim (\mathbf{3^*}, \mathbf{1}, \tfrac{4}{3}) \,, \quad \phi \sim (\mathbf{3}, \mathbf{2}, \tfrac{7}{6}) \,, \quad \eta \sim (\mathbf{3}, \mathbf{1}, \tfrac{5}{3}) \,.
\end{equation}
The colour factor in \cref{eq:mf1,eq:yeff1} is then $n_3=2$, while $n_2=1$. The hypercharge $Y_\psi$ has been chosen such that $\phi$ is a leptoquark, with the possible additional interactions
\begin{equation}
    \mathcal{L}_\text{LQ} = y_{Lu}^{ij} \bar{L^i} \phi^\dagger u_R^j + y_{Qe}^{ij} \bar{Q^i} \phi e_R^j + \text{h.c.} \,,
\end{equation}
where $i,j$ are flavour indices. The couplings $y_{Lu}$, $y_{Qe}$ break $U(1)_X$ and allow the exotic states to decay into the SM. More precisely, $X$ and $B$ are broken to the combination $X-3B/2$.  We assume these couplings are present so there are no constraints from coloured or charged relics.

Care should be taken when comparing the expression for the $b$-quark mass with its measured value, since higher-order QCD effects are numerically significant. We therefore match \cref{eq:mf1} onto the $\overline{\text{MS}}$ quark mass in the SM, taking the matching scale to be the geometric mean of $m_1$, $m_2$ and $m_\psi$. The SM running is performed using the 5-loop results implemented in {\tt RunDec}~\cite{hep-ph/0004189,1703.03751}. To demonstrate the importance of the renormalisation group evolution, at a matching scale of $10^4\,$GeV the $\overline{\text{MS}}$ $b$-quark mass is roughly a factor of 2 smaller than its value at $m_b(m_b)=4.18\,$GeV.

%--------------------------------------------------------
\begin{figure}[ht]
    \centering
    \includegraphics[height=0.45\textwidth]{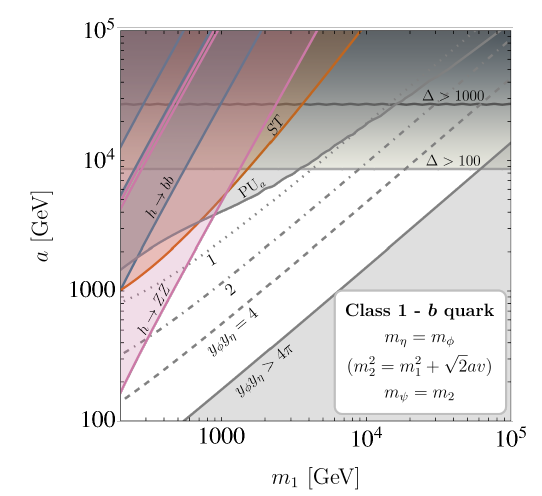}
    \includegraphics[height=0.45\textwidth]{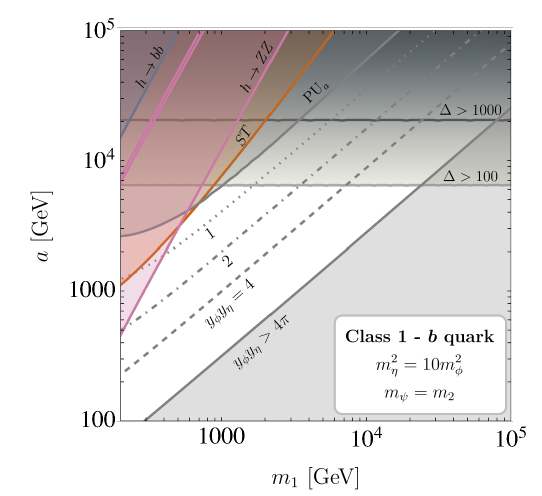}
	\caption{Theoretical (grey) and current $2\sigma$ experimental constraints (coloured) on the  Class~1 $b$-quark benchmark model.}
	\label{fig:1b}
\end{figure}
%--------------------------------------------------------

In \cref{fig:1b} we present the theoretical and current experimental constraints on the same two slices of parameter space considered in \cref{sec:class1-tau}.  We see that the required product of Yukawas is smaller here, due to the colour factor entering the radiative mass ($m_b \approx m_\tau$ at the matching scale).  The naturalness bound is stronger due to the multiplicity of states running in the loop which contributes to the Higgs mass, seen in the factor of $d_3(\phi)=3$ in \cref{eq:class1-fine-tuning}, and the perturbative unitarity constraint on the trilinear coupling is slightly stronger.  Again, we see that the $b$-quark mass can be generated radiatively with order-one Yukawas.

On the experimental side, we see that the $2\sigma$ constraint from the Higgs to $\bar{b}b$ signal strength is weaker than the analogous $\tau$ limit in the Class~1 tau benchmark model. While the deviation in the $h\bar{b}b$ coupling here is identical to the deviation of the $h\bar{\tau}\tau$ coupling in the previous benchmark model and the experimental sensitivities are comparable, there is an approximate cancellation in the $h \to \bar{b}b$ signal strength due to an increase in the width of the Higgs, since the Higgs decays dominantly to $\bar{b}b$.  This cancellation does not occur in the tau benchmark model since the Higgs branching ratio to taus is much smaller.  

An important observation is that in these models the new contribution to the $hgg$ coupling interferes destructively with the SM contribution.  This means that the $hgg$ coupling can be either smaller or larger than the SM coupling in different regions of parameter space.  An interplay between the $h\bar{b}b$ coupling and the $hgg$ coupling then results in an unconstrained strip at large trilinear coupling.  While this region is theoretically disallowed due to the perturbative unitarity constraint on the trilinear coupling, it is interesting as it demonstrates how correlations between the modified Higgs couplings in these models can impact the signal strength measurements.

For Higgs measurements in the SM gauge boson channels, the $h \to ZZ$ signal strength gives a stronger constraint than $h \to \gamma\gamma$.  In the relevant regions of parameter space, the gluon-fusion production cross-section is suppressed, while the diphoton branching ratio is enhanced. This leads to a partial cancellation of the new contributions to the diphoton signal strength in this model, resulting in a weaker constraint than $h \to ZZ$.

Finally, the constraint from the $S$ and $T$ parameters is slightly stronger in this model, due to the colour factor $d_3(\phi)=3$ entering the self-energy diagrams (reflected in the $d_3(\phi)$ factor seen in \cref{eq:class1-S}).

%--------------------------------------------------------
\begin{figure}[ht]
    \centering
    \includegraphics[height=0.45\textwidth]{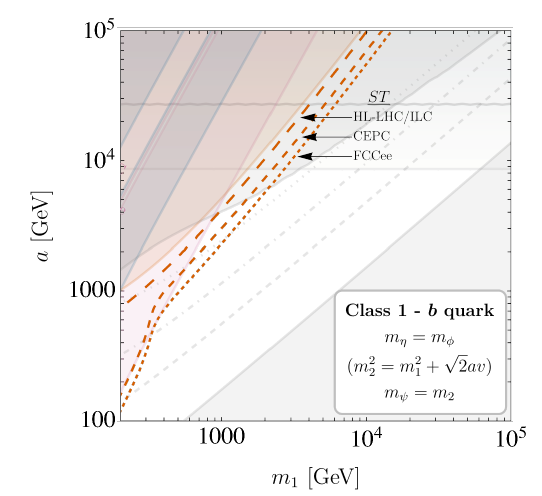}
    \includegraphics[height=0.45\textwidth]{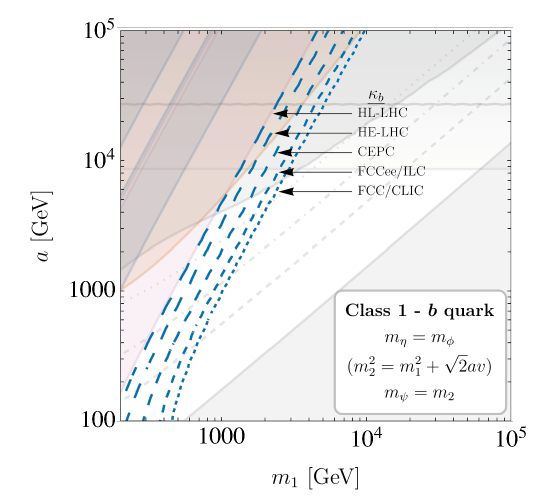}
    \\
    \includegraphics[height=0.45\textwidth]{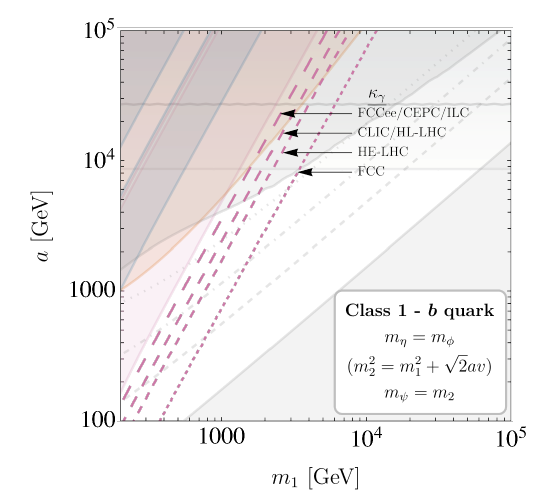}
    \includegraphics[height=0.45\textwidth]{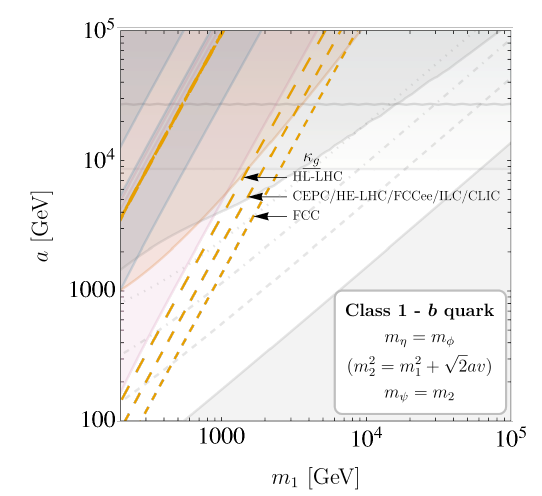}
	\caption{Projected $2\sigma$ experimental sensitivities on the Class~1 $b$-quark benchmark model at future colliders for the left hand slice in \cref{fig:1b}.}
	\label{fig:1b-FC}
\end{figure}
%--------------------------------------------------------

In \cref{fig:1b-FC} we present the projected sensitivity of future experiments on the same parameter space as the left-hand panel of \cref{fig:1b}.  We see that the improvements in $S$ and $T$ are very comparable to the Class~1 $\tau$ case, except at small trilinear coupling and exotic masses. In most of the plot the sensitivity is driven by the $T$ parameter, which scales as $a^4$ compared to $a^2$ for $S$ (see \cref{eq:class1-S}); however, in this bottom-left region the $S$ parameter has the dominant deviation.
Future measurements of the $Z\bar{b}b$ couplings are not sensitive to this model, reflecting the modest improvement in precision for these couplings expected at future colliders.  
The future $\kappa_b$ and $\kappa_\gamma$ sensitivities are the result of a global fit, so do not suffer from the cancellations discussed above for the current $\bar{b}b$ and diphoton signal strength constraints.  The future sensitivity from $\kappa_\gamma$ is now competitive with $\kappa_b$ due to the larger hypercharges of the scalars in this model compared to the Class~1 $\tau$ benchmark model.  Finally, although there is sensitivity from $\kappa_g$, due to the coloured scalars, it is less sensitive than $\kappa_b$ and $\kappa_\gamma$.

Last, we comment on collider searches for the exotic states. Since these are coloured they can be pair produced at the LHC purely via QCD interactions. Furthermore, ATLAS and CMS have dedicated searches for scalar leptoquarks. If the scalars decay predominantly into first (second) generation leptons, the current bound is $m_1>1.8\,(1.7)$\,TeV~\cite{2006.05872}, and  $m_1>1.5$\,TeV for decays involving the top quark~\cite{2010.02098}. If they decay only into third generation fermions the limits are significantly weaker:~$m_1>920$\,GeV~\cite{1902.08103}. Note that these bounds assume that either $y_{Lu}$ or $y_{Qe}$ is sufficiently large that the decays are prompt. If these couplings are $\mathcal{O}(1)$, then single production may also provide competitive bounds.

%--------------------------------------------------------
\subsection{\texorpdfstring{Class~1 Combined $b+\tau$ Benchmark Model}{Class~1 Combined b + tau Benchmark Model}}
\label{sec:class1-pheno-combined}
%--------------------------------------------------------

Finally for Class~1, we consider a combined benchmark model which can radiatively generate both the $\tau$-lepton and $b$-quark masses.  As discussed in \cref{sec:class1-classification-combined}, there are fewer choices for the quantum numbers in these models and the hypercharges are fixed.  As our benchmark we take the minimal model containing the following fields
\begin{equation}
    \psi_{L,R} \sim (\mathbf{3^*}, \mathbf{1}, \tfrac{1}{3}) \,, \quad \phi \sim (\mathbf{3}, \mathbf{2}, \tfrac{1}{6}) \,, \quad \eta \sim (\mathbf{3}, \mathbf{1}, \tfrac{2}{3}) \,.
\end{equation}
These fields have the same $SU(3)_C \times SU(2)_L$ quantum numbers as the $b$-quark model of the previous section, but with smaller hypercharges.  The group theory factors for this model are $n_2^\tau = 1$, $n_3^\tau = 3$, $n_2^b = 1$ and $n_3^b = 2$.  As discussed in \cref{sec:class1-classification-combined}, there is no $U(1)_X$ symmetry in the combined models but there remains an exotic $\mathcal{Z}_2$.  There is no obviously viable dark matter candidate in this model, but there are four possible interactions with SM fields which can break the $\mathcal{Z}_2$ symmetry:~$\overline{d}_R^i L_L^j \phi$, $\overline{(d_R^i)^c} d_R^j \eta$, $\bar{Q} H \psi_L^c$, and $\overline{d}_R^i \psi_R^c$. Note that all of these terms also break the generalised baryon and lepton number symmetries to the combination $B+L$. Hence, $\mathcal{Z}_2$ breaking leads to proton decay. We estimate the proton lifetime below, and demonstrate that the experimental bounds can be satisfied.

%--------------------------------------------------------
\begin{figure}[ht]
    \centering
    \includegraphics[height=0.45\textwidth]{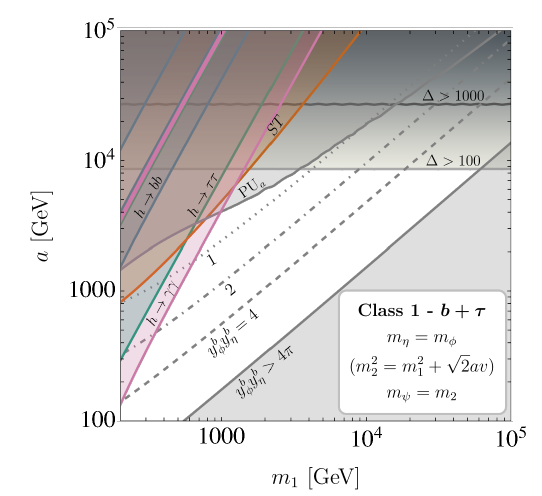}
    \includegraphics[height=0.45\textwidth]{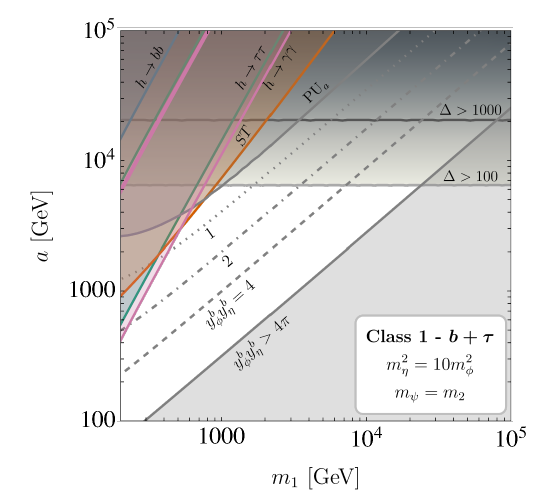}
	\caption{Theoretical (grey) and current $2\sigma$ experimental constraints (coloured) on the Class~1 combined $b+\tau$ benchmark model.}
	\label{fig:1comb}
\end{figure}
%--------------------------------------------------------

In \cref{fig:1comb} we show the existing theoretical constraints on the same slices of parameter space as seen previously.  The perturbativity constraints on the Yukawa couplings $y_\phi^b y_\eta^b$ are identical to those in \cref{fig:1b}.  While there is a perturbativity constraint on $y_\phi^\tau y_\eta^\tau$, this is weaker than on $y_\phi^b y_\eta^b$ and we do not show it.  The perturbative unitarity constraints on the trilinear coupling and the fine-tuning are identical to the Class~1 $b$-quark benchmark model, since the relevant group theory factors are the same.

The current $2\sigma$ experimental constraints are shown in colour.  The Higgs to $\tau^+\tau^-$ signal strength constraint is stronger than in \cref{fig:1tau}.  While this signal strength is greater than 1 in the Class~1 $\tau$ model, here it is less than 1 due to a suppression in the gluon-fusion production cross-section and an increased total width (as the dominant partial width, $\bar{b}b$, is increased).  At larger values of the trilinear coupling the gluon-fusion production cross-section becomes enhanced, and the transition between these regimes leads to a narrow gap in this and other exclusions.  In this model the Higgs to diphoton signal strength provides a very similar constraint to $ZZ$, but we only show the former as it is slightly stronger.  The constraints from the electroweak $S$ and $T$ parameters are identical to \cref{fig:1b}.  Current measurements of the $Z$ couplings to third generation fermions do not provide any constraints on the parameter space.  Overall, we see that there is viable parameter space in both slices that can radiatively generate both the $b$-quark and $\tau$-lepton masses with exotic Yukawas of order one.

%--------------------------------------------------------
\begin{figure}[t]
    \centering
    \includegraphics[height=0.45\textwidth]{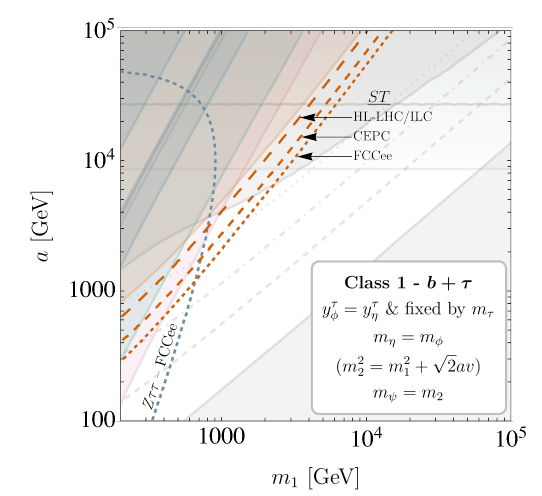}
    \includegraphics[height=0.45\textwidth]{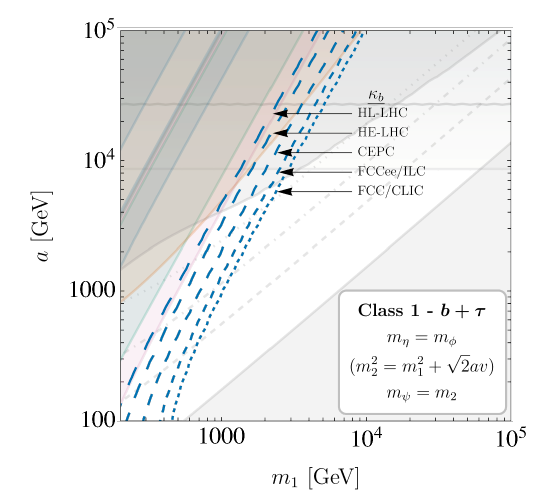}
    \\
    \includegraphics[height=0.45\textwidth]{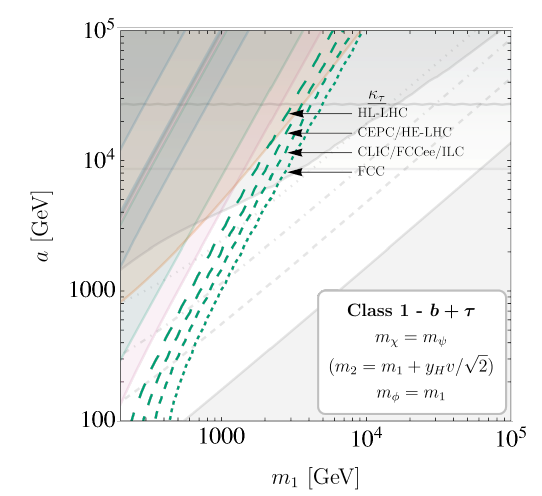}
    \includegraphics[height=0.45\textwidth]{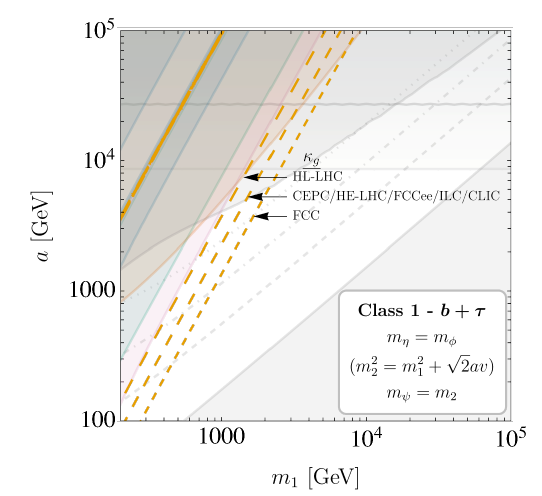}
	\caption{Projected $2\sigma$ experimental sensitivities on the Class~1 combined $b+\tau$ benchmark model at future colliders for the left hand slice in \cref{fig:1comb}.}
	\label{fig:1comb-FC}
\end{figure}
%--------------------------------------------------------

In the top-left panel of \cref{fig:1comb-FC} we see that the sensitivities of future measurements of $S$ and $T$ and the $Z\bar{\tau}\tau$ couplings are very similar to those seen in \cref{fig:1tau-FC}.  Again, the expected improvements in measurements of the $Z\bar{b}b$ couplings are not large enough to provide sensitivity.  In the top-right and bottom-left panels we see the expected future sensitivities from $\kappa_b$ and $\kappa_\tau$, which are identical to those seen in \cref{fig:1tau-FC,fig:1b-FC}.  Even though the quantum numbers in the models differ, all group theory factors cancel out for these observables when the models are required to reproduce the SM fermion masses.  In the bottom-right panel we see the improvement due to future sensitivity from $\kappa_g$, which is identical to that seen in \cref{fig:1b-FC} as the $SU(3)_C$ representations of the exotics are the same in these models.

Since all of the exotic fields are charged under $SU(3)_C$, there is no dark matter candidate in this model. Even if the lightest QCD bound state containing an exotic is electrically neutral (for example, the meson $\psi d$), it will still have residual strong interactions with other hadrons. As discussed above, there are strong bounds on such cosmological relics. In some regions of parameter space the abundance may be low enough to satisfy these bounds, but a straightforward way to evade them is to extend the exotic sector to include a new $SU(3)_C$-singlet field that can be dark matter. 

Alternatively, we can retain the minimal field content but break the $\mathcal{Z}_2$ symmetry so that the exotics decay. As noted above, however, this leads to proton decay which introduces stringent bounds on any $\mathcal{Z}_2$-breaking couplings. We consider the term\footnote{This term would in any case be radiatively generated by the other $\mathcal{Z}_2$-breaking couplings. It also softly breaks both $S_\psi$ and $S_a$; however, since proton decay requires the coupling to be extremely small, its effect on fermion mass generation is completely negligible.} $\mu_\slashed{\mathcal{Z}_2} \bar{d_R^i} \psi_R^c$, since it leads to proton decay via the lowest possible dimension operator (dimension-7, as $B+L$ is preserved) through tree-level scalar exchange. Given we are remaining agnostic about the mass generation mechanism for the first- and second-family fermions, we consider the analogous process to proton decay in the third-family. This gives a conservative estimate of the proton lifetime, since the true proton decay amplitude may contain additional suppression by either off-diagonal CKM elements or $(m_p/m_W)^2$. The decay rate is then approximately

\begin{equation}
    \Gamma_p \sim \frac{|y_\phi^b y_\eta^\tau|^2}{512\pi^3} \left(\frac{\mu_\slashed{\mathcal{Z}_2}}{m_\psi}\right)^4 \bigg(\frac{av}{\sqrt{2}m_\eta^2}\bigg)^2\frac{m_p^5}{m_\phi^4} \,.
\end{equation}
To compare this with the bound on the proton lifetime we need to consider how small the coupling $\mu_\slashed{\mathcal{Z}_2}$ can be. A naive bound is obtained by requiring that the exotics decay prior to BBN. Requiring a lifetime shorter than 1 second leads to the bound
\begin{equation}
    \left(\frac{\mu_\slashed{\mathcal{Z}_2}}{m_\psi}\right)^2 \gtrsim 10^{-25} \left(\frac{\text{TeV}}{M} \right) \,,
\end{equation}
where $M$ is the mass of the decaying exotic (the decay rate also depends on one of $y_\phi^\tau$, $y_\eta^b$ or $g_s$, which are all $\mathcal{O}(1)$). Saturating this bound, and setting all new physics scales equal to $M$, leads to a proton lifetime of
\begin{equation} \label{eq:class1-proton-lifetime}
  \tau_p \sim 10^{36} \left(\frac{M}{\text{TeV}}\right)^8 \text{yr} \,.
\end{equation}
Experimental bounds on the proton lifetime range from $6\times10^{29}$\, years for the inclusive mode $p \to e^+ +  X$~\cite{10.1103/PhysRevLett.43.907} up to $10^{34}$ years for $p \to e^+ \pi^0$ \cite{1705.07221}. It is clear from \cref{eq:class1-proton-lifetime} that even the stronger bound is satisfied for exotic masses above the weak scale. However, larger values of $\mu_\slashed{\mathcal{Z}_2}$ will lead to proton decay that could be observable at current and future experiments, such as Hyper-Kamiokande. The small values of $\mu_\slashed{\mathcal{Z}_2}$ above are somewhat in conflict with the motivation for radiative models, which seek to explain hierarchies in fermion Yukawas; however, as noted above these proton decay bounds apply only to the minimal model and can be easily avoided by extending the model, for example by including a dark matter candidate.

Finally, since the exotics must be charged under $SU(3)_C$ in the combined models, LHC searches also provide important constraints. Given the upper bound on the $\mathcal{Z}_2$-breaking couplings from proton decay, the lightest exotic is expected to be collider-stable in almost all of the viable parameter space. A long-lived coloured state will hadronise to form $R$-hadrons, which have been targeted in dedicated LHC searches. If $\tilde{\eta}$ or $\tilde{\phi}$ is the lightest exotic, then one can immediately impose the bounds from existing SUSY searches for long-lived stops or sbottoms, leading to $m_{1,2}\gtrsim1.3$\,TeV~\cite{1902.01636}. Recasting these searches should lead to a similar bound on $m_\psi$ in the case where $\psi$ is the lightest exotic.

%=============================================================================
\section{Class~2 -- Model Classification}
\label{sec:class2-classification}
%=============================================================================

We now move on to discuss the second class of models, which radiatively generate Yukawa couplings via the right diagram in \cref{fig:feyn}.  These models contain two new vector-like fermions and a new scalar.  As in Class~1, we classify models which can generate the Yukawa coupling for the $\tau$-lepton, the $b$-quark and both the $\tau$-lepton and the $b$-quark at the same time.  We again highlight the models which contain only singlets and fundamentals of the SM gauge group, from which we take the benchmark models in \cref{sec:class2-pheno}.

%==============================================================================
\subsection{\texorpdfstring{Class~2 $\tau$-lepton Model Classification}{Class~2 tau-lepton Model Classification}}
%==============================================================================

For a model to generate an effective Yukawa coupling for the $\tau$-lepton via the right diagram in \cref{fig:feyn}, its Lagrangian requires the following terms,
\begin{equation}
     \mathcal{L}_\tau \supset\, -y_\psi \overline{L}_L \phi^\dagger \psi_R - y_\chi \overline{\chi}_L \phi \tau_R - y_H \overline{\psi}_L H \chi_R - m_\psi \overline{\psi}_L \psi_R - m_\chi \overline{\chi}_L \chi_R + \mathrm{h.c.} \,,
     \label{eq:Lagtau2}
\end{equation}
where $L_L$ is the third-family left-handed lepton doublet, $\tau_R$ is the right-handed $\tau$ lepton, $H$ is the Higgs doublet, and $\psi$, $\chi$ and $\phi$ are exotic fields. We can take all couplings to be positive without loss of generality. The possible $SU(3)_C$ and $SU(2)_L$ representations of the new particles are given in \cref{tab:class2-DynkinLabels} (left). 

%============================================================
\begin{table}[ht]
    \begin{subtable}[h]{0.35\textwidth}
        \begin{tabular}{@{\hspace{1em}} c @{\hspace{2em}} c @{\hspace{2em}} c @{\hspace{1em}}}
            \toprule
            & $SU(3)_C$ & $SU(2)_L$ \\ 
            \midrule
            $\psi$ & $(a,b)$ & $(c)$ \\
            $\chi$ & $(a,b)$ & $(|c \pm 1|)$ \\
            $\phi$ & $(a,b)$ & $(|c\pm 1|)$ \\
            \bottomrule
        \end{tabular}
        \caption{$\tau$-lepton}
        \label{tab:class2tau-DynkinLabels}
    \end{subtable}
    \begin{subtable}[h]{0.55\textwidth}
        \begin{tabular}{@{\hspace{1em}} c @{\hspace{2em}} c c  c @{\hspace{2em}} c @{\hspace{1em}}}
            \toprule
            & \multicolumn{3}{c}{$SU(3)_C$} \hspace{1em} & $SU(2)_L$ \\ 
            & I & II & III & \\  
            \midrule
            $\psi$ & $(a,b)$ & $(a+1,b)$ & $(a,b+1)$ & $(c)$ \\
            $\chi$ & $(a,b)$  & $(a+1,b)$ & $(a,b+1)$ & $(|c \pm 1|)$ \\
            $\phi$ & $(a,b+1)$ & $(a,b)$ & $(a+1,b)$ & $(|c \pm 1|)$ \\
            \bottomrule
        \end{tabular}
        \caption{$b$-quark}
        \label{tab:class2b-DynkinLabels}
    \end{subtable}
    \caption{Dynkin labels giving the allowed colour and weak-isospin assignments of the new fields in Class~2 for the $\tau$-lepton (left) and $b$-quark (right). For the $b$-quark models there are three possible patterns for the $SU(3)_C$ assignments.}
    \label{tab:class2-DynkinLabels}
\end{table}
%============================================================

%==============================================================
\begin{table}[ht]
    \begin{tabular}{@{\hspace{1em}} c @{\hspace{2em}} c @{\hspace{2em}} c @{\hspace{2em}} c @{\hspace{2em}} c @{\hspace{2em}} c @{\hspace{2em}} c @{\hspace{2em}} c @{\hspace{2em}} c @{\hspace{1em}}}
        \toprule
        & $L_L$ & $\tau_R$ & $H$ & $\psi_L$ & $\psi_R$ & $\chi_L$ & $\chi_R$ & $\phi$ \\
        \midrule
        $Y$ & $-\frac{1}{2}$ & $-1$ & $\frac{1}{2}$ & $Y_\psi$ & $Y_\psi$ & $Y_\psi\! -\! \frac{1}{2}$ & $Y_\psi\! -\! \frac{1}{2}$ & $Y_\psi\! +\! \frac{1}{2}$ \\
        $L$ & $1$ & $1$ & $0$ & $0$ & $0$ & $0$ & $0$ & $-1$\\
        $X$ & $0$ & $0$ & $0$ & $1$ & $1$ & $1$ & $1$ & $1$ \\
        \midrule
        $S_\chi$ & $0$ & $1$ & $0$ & $0$ & $0$ & $1$ & $0$ & $0$ \\
        $S_\psi$ & $0$ & $1$ & $0$ & $0$ & $-1$ & $0$ & $0$ & $-1$ \\
        \midrule
        $SU(3)_C\! \times\! SU(2)_L$ & $(1,2)$ & $(1,1)$ & $(1,2)$ & $(1,1)$ & $(1,1)$ & $(1,2)$ & $(1,2)$ & $(1,2)$ \\
        &&&& $(1,2)$ & $(1,2)$ & $(1,1)$ & $(1,1)$ & $(1,1)$ \\
        &&&& $(3,1)$ & $(3,1)$ & $(3,2)$ & $(3,2)$ & $(3,2)$ \\
        &&&& $(3,2)$ & $(3,2)$ & $(3,1)$ & $(3,1)$ & $(3,1)$ \\
        &&&& $(3^*,1)$ & $(3^*,1)$ & $(3^*,2) $ & $(3^*,2)$ & $(3^*,1)$ \\
        &&&& $(3^*,2)$ & $(3^*,2)$ & $(3^*,1)$ & $(3^*,1)$ & $(3^*,1)$ \\
        \bottomrule
    \end{tabular}
    \caption{Quantum numbers of fields for Class~2 $\tau$-lepton models. The last six lines list the models containing only singlets and fundamentals under the SM gauge group.}
    \label{tab:class2-tau}
\end{table}
%============================================================================

The above Lagrangian possesses three exact $U(1)$ symmetries:~hypercharge ($Y$), lepton number ($L$) and exotic particle number ($X$). The charges of the fields under these symmetries are given in \cref{tab:class2-tau}.  The hypercharges of the exotic fields contain a free parameter, which we take to be $Y_\psi$.  We choose to assign $L=0$ to $\psi$ and $\chi$; $\phi$ then has $L=-1$. We normalise $X=1$ for the exotics.  Again, $U(1)_X$ acts to stabilise the exotic particles and if unbroken may lead to a dark matter candidate.  

There are two softly broken symmetries which forbid the $\tau$-lepton Yukawa coupling with the Higgs. The mass terms for $\psi$ and $\chi$ softly break the symmetries $S_\psi$ and $S_\chi$, respectively, and the charges are given in \cref{tab:class2-tau}.  We also list the minimal models containing only singlets and fundamentals under the SM gauge group.

%==============================================================================
\subsection{\texorpdfstring{Class~2 $b$-quark Model Classification}{Class~2 b-quark Model Classification}}
\label{sec:class2-classification-combined}
%==============================================================================

For the $b$-quark case, the Lagrangian contains the following terms,
\begin{equation}
    \mathcal{L}_b \supset\, -y_t \overline{Q}_L \tilde{H} t_R - y_\psi \overline{Q}_L \phi^\dagger \psi_R - y_\chi \overline{\chi}_L \phi b_R - y_H \overline{\psi}_L H \chi_R - m_\psi \overline{\psi}_L \psi_R - m_\chi \overline{\chi}_L \chi_R + \mathrm{h.c.}
    \label{eq:Lagb2}
\end{equation}
The possible $SU(3)_C$ and $SU(2)_L$ quantum numbers are given in \cref{tab:class2-DynkinLabels} (right). The three unbroken $U(1)$ symmetries are hypercharge ($Y$), baryon number ($B$) and exotic particle number ($X$), with the quantum numbers of the fields shown in \cref{tab:class2-bquark}. The $b$-quark Yukawa coupling with the Higgs is forbidden by the $U(1)$ symmetries $S_\psi$ and $S_\chi$, which are again softly broken by the $\psi$ and $\chi$ masses respectively. \Cref{tab:class2-bquark} also lists the six models which contain only singlets or fundamentals of $SU(3)_C$ and $SU(2)_L$.

%============================================================
\begin{table}[ht]
    \begin{tabular}{@{\hspace{0.8em}} c @{\hspace{1.6em}} c @{\hspace{1.6em}} c @{\hspace{1.6em}} c @{\hspace{1.6em}} c @{\hspace{1.6em}} c @{\hspace{1.6em}} c @{\hspace{1.6em}} c @{\hspace{1.6em}} c @{\hspace{1.6em}} c @{\hspace{0.8em}}}
        \toprule
        & $Q_L$ & $t_R$ & $b_R$ & $H$ & $\psi_L$ & $\psi_R$ & $\chi_L$ & $\chi_R$ & $\phi$ \\
        \midrule
        $Y$ & $\frac{1}{6}$ & $\frac{2}{3}$ & -$\frac{1}{3}$ & $\frac{1}{2}$ & $Y_\psi$ & $Y_\psi$& $Y_\psi\! -\! \frac{1}{2}$ & $Y_\psi\! -\! \frac{1}{2}$ & $Y_\psi\! -\! \frac{1}{6}$ \\
        $B$ & $\frac{1}{3}$ & $\frac{1}{3}$ & $\frac{1}{3}$ & $0$ & $0$ & $0$ & $0$ & $0$ & $-\frac{1}{3}$\\
        $X$ & $0$ & $0$ & $0$ & $0$ & $1$ & $1$ & $1$ & $1$ & $1$ \\
        \midrule
        $S_\chi$ & $0$ & $0$ & $1$ & $0$ & $0$ & $0$ & $1$ & $0$ & $0$ \\
        $S_\psi$ & $0$ & $0$ & $1$ & $0$ & $0$ & $-1$ & $0$ & $0$ & $-1$\\
        \midrule
        $SU(3)_C\! \times\! SU(2)_L$ & $(3,2)$ & $(3,1)$ & $(3,1)$ & $(1,2)$ & $(1,1)$ & $(1,1)$ & $(1,2)$ & $(1,2)$ & $(3^*,2)$ \\
        &&&&& $(1,2)$ & $(1,2)$ & $(1,1)$ & $(1,1)$ & $(3^*,1)$ \\
        &&&&& $(3,1)$ & $(3,1)$ & $(3,2)$ & $(3,2)$ & $(1,2)$ \\
        &&&&& $(3,2)$ & $(3,2)$ & $(3,1)$ & $(3,1)$ & $(1,1)$ \\
        &&&&& $(3^*,1)$ & $(3^*,1)$ & $(3^*,2) $ & $(3^*,2)$ & $(3,2)$ \\
        &&&&& $(3^*,2)$ & $(3^*,2)$ & $(3^*,1)$ & $(3^*,1)$ & $(3,1)$ \\
        \bottomrule
    \end{tabular}
    \caption{Quantum numbers of fields for Class~2 $b$-quark models. The last six lines list the models containing only singlets and fundamentals under the SM gauge group.}
    \label{tab:class2-bquark}
\end{table}
%===============================================================

%==============================================================================
\subsection{\texorpdfstring{Class~2 Combined $b+\tau$ Model Classification}{Class~2 Combined b+tau Model Classification}}
%==============================================================================

Finally, we consider models that can radiatively generate both the $b$-quark and $\tau$-lepton masses simultaneously. From \cref{tab:class2-bquark,tab:class2-tau}, it is immediately clear that the fermions $\psi$ and $\chi$ can be common to both the quark and lepton sectors.  This identification yields the most economical models. There are then two possibilities for the scalars. First, there is the option to simply introduce two separate fields $\phi_b$ and $\phi_\tau$ in the $b$ and $\tau$ sectors, respectively. A more interesting possibility is obtained by noticing that one can identify $\phi_\tau = \phi_b^*$. This is possible only if $Y_\psi=-1/6$ and if the $SU(3)_C$ representations satisfy $\psi,\chi \sim (a,a+1)$ and $\phi \sim (a+1,a)$, pattern III in \cref{tab:class2-DynkinLabels} (right) with $a=b$.

Focusing on the models in which all the exotic fields are common to both sectors, the relevant Lagrangian is
\begin{align} \label{eq:Lagcomb2}
    \mathcal{L} \supset &\, -y_t\, \overline{Q}_L \tilde{H} t_R - y_\psi^b\, \overline{Q}_L \phi^\dagger \psi_R - y_\chi^b\, \overline{\chi}_L \phi b_R - y_\psi^\tau\, \overline{L}_L \phi \psi_R - y_\chi^\tau\, \overline{\chi}_L \phi^\dagger \tau_R \notag\\
    &\,- y_H\, \overline{\psi}_L H \chi_R - m_\psi \overline{\psi}_L \psi_R - m_\chi \overline{\chi}_L \chi_R + \mathrm{h.c.}
\end{align}
The above Lagrangian exhibits three unbroken $U(1)$ symmetries:~hypercharge, and generalised baryon and lepton numbers. There is also a $\mathcal{Z}_2$ parity under which the exotic fields are odd. Two softly broken $U(1)$ symmetries, again denoted by $S_\chi$ and $S_\psi$, forbid both the $b$ and $\tau$ Yukawa couplings. The charges of the fields under these symmetries and the minimal models are given in \cref{tab:class2-combined}.

%==============================================================
\begin{table}[ht]
    \begin{tabular}{@{\hspace{1em}} c @{\hspace{1.2em}} c @{\hspace{1.2em}} c @{\hspace{1.2em}} c @{\hspace{1.2em}} c @{\hspace{1.2em}} c @{\hspace{1.2em}} c @{\hspace{1.2em}} c @{\hspace{1.2em}} c @{\hspace{1.2em}} c @{\hspace{1.2em}} c @{\hspace{1em}}}
        \toprule
        & $Q_L$ & $b_R$ & $L_L$ & $\tau_R$ & $H$ & $\psi_L$ & $\psi_R$ & $\chi_L$ & $\chi_R$ & $\phi$ \\
        \midrule
        $Y$ & $\frac{1}{6}$ & $-\frac{1}{3}$ & $-\frac{1}{2}$ & $-1$ & $\frac{1}{2}$ & $-\frac{1}{6}$ & $-\frac{1}{6}$ & $-\frac{2}{3}$ & $-\frac{2}{3}$ & $-\frac{1}{3}$ \\
        $B$ & $\frac{1}{3}$ & $\frac{1}{3}$ & $0$ & $0$ & $0$ & $\frac{1}{6}$ & $\frac{1}{6}$ & $\frac{1}{6}$ & $\frac{1}{6}$ & $-\frac{1}{6}$\\
        $L$ & $0$ & $0$ & $1$ & $1$ & $0$ & $\frac{1}{2}$ & $\frac{1}{2}$ & $\frac{1}{2}$ & $\frac{1}{2}$ & $\frac{1}{2}$ \\
        $\mathcal{Z}_2$ & 1 & 1 & 1 & 1 & 1 & $-1$ & $-1$ & $-1$ & $-1$ & $-1$ \\
        \midrule
        $S_\chi$ & $0$ & $1$ & $0$ & $1$ & $0$ & $0$ & $0$ & $1$ & $0$ & $0$ \\
        $S_\psi$ & $0$ & $1$ & $0$ & $1$ & $0$ & $1$ & $0$ & $1$ & $1$ & $0$ \\
        \midrule
        $SU(3)_C\! \times\! SU(2)_L$ & $(3,2)$ & $(3,1)$ & $(1,2)$ & $(1,1)$ & $(1,2)$ & $(3^*,2)$ & $(3^*,2)$ & $(3^*,1)$ & $(3^*,1)$ & $(3,1)$ \\
        &&&&&& $(3^*,1)$ & $(3^*,1)$ & $(3^*,2)$ & $(3^*,2)$ & $(3,2)$ \\
        \bottomrule
    \end{tabular}
    \caption{Quantum numbers of the fields for the minimal Class~2 combined models. The last two lines list the models containing only singlets and fundamentals under the SM gauge group.}
    \label{tab:class2-combined}
\end{table}
%============================================================================

%==============================================================================
\section{Class~2 - Radiative Mass Generation}
\label{sec:class2-mass}
%==============================================================================

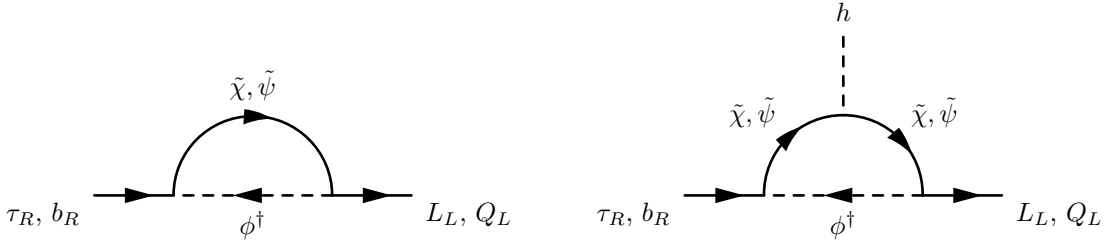
\begin{figure}[t]
  \begin{center}
	% !TEX root = ../paper.tex
\noindent\makebox[\columnwidth][c]{
\begin{tabular}{ccc}
\begin{fmffile}{feyngraph-4a}
\begin{fmfgraph*}(120,60)
    \fmfstraight
    \fmfleft{l1,l2}
    \fmfright{r1,r2}
    \fmf{fermion}{l1,v1}
    \fmf{dashes_arrow,tension=0.5,label=$\phi^\dagger$,label.side=left}{v2,v1}
    \fmf{fermion}{v2,r1}
    \fmf{phantom}{l2,v4,r2}
    \fmffreeze
    \fmf{fermion,left,label=$\tilde{\chi},,\tilde{\psi}$}{v1,v2}
    \fmflabel{$\tau_R$, $b_R$}{l1}
    \fmflabel{$L_L$, $Q_L$}{r1}
  \end{fmfgraph*}
\end{fmffile}
& \hspace{3cm} &
\begin{fmffile}{feyngraph-4b}
\begin{fmfgraph*}(120,60)
    \fmfstraight
    \fmfleft{l1,l2}
    \fmfright{r1,r2}
    \fmf{fermion}{l1,v1}
    \fmf{dashes_arrow,tension=0.5,label=$\phi^\dagger$,label.side=left}{v2,v1}
    \fmf{fermion}{v2,r1}
    \fmf{phantom}{l2,v4,r2}
    \fmffreeze
    \fmf{fermion,left=0.42,label=$\tilde{\chi},,\tilde{\psi}$}{v1,v3}
    \fmf{fermion,left=0.42,label=$\tilde{\chi},,\tilde{\psi}$}{v3,v2}
    \fmf{dashes,tension=2.05}{v4,v3}
    \fmflabel{$\tau_R$, $b_R$}{l1}
    \fmflabel{$h$}{v4}
    \fmflabel{$L_L$, $Q_L$}{r1}
  \end{fmfgraph*}
\end{fmffile}
\end{tabular}
}
  \end{center}
	\caption{One-loop diagrams that generate effective $\tau$-lepton and $b$-quark masses and effective Yukawa couplings after electroweak symmetry breaking in the Class~2 models.}
	\label{fig:feyn-class-2-after-ewsb}
\end{figure}

In this class of models, electroweak symmetry breaking leads to a mixing of the fermions $\psi$ and $\chi$ via the $y_H$-terms in \cref{eq:Lagb2,eq:Lagtau2,eq:Lagcomb2}.  The diagrams which generate the fermion mass and effective Yukawa coupling are shown in the left and right sides of \cref{fig:feyn-class-2-after-ewsb}.  As discussed for Class~1 models, there are corrections to the SM prediction for the fermion coupling to the physical Higgs boson at $\mathcal{O}(v^2/M^2)$, where $M$ is the scale of new physics. 

%========================================================
%=========================================================

There are $n_2\equiv\min(d_2(\psi),d_2(\chi))$ pairs of mixed states, where $d_2$ is the dimension of the $SU(2)_L$ representation.  As in Class~1, the larger $SU(2)_L$ multiplet contains one state that does not mix. The mass matrix of the states that mix is diagonalised by the following transformation
\begin{equation}
    \begin{pmatrix}
        \tilde{\psi}_L \\
        \tilde{\chi}_L
    \end{pmatrix}
    =
    \begin{pmatrix}
        \cos\theta_L & -\sin\theta_L \\
        \sin\theta_L & \cos\theta_L
    \end{pmatrix}
    \begin{pmatrix}
        \psi_L \\
        \chi_L
    \end{pmatrix}
    \,, \qquad
    \begin{pmatrix}
        \tilde{\psi}_R \\
        \tilde{\chi}_R
    \end{pmatrix}
    =
    \begin{pmatrix}
        \cos\theta_R & -\sin\theta_R \\
        \sin\theta_R & \cos\theta_R
    \end{pmatrix}
    \begin{pmatrix}
        \psi_R \\
        \chi_R
    \end{pmatrix}
    \,,
\end{equation}
where $\tilde{\psi}_{L,R}$, $\tilde{\chi}_{L,R}$ are the mass basis fields and $\theta_L,\theta_R \in [0,\tfrac{\pi}{2}]$. The corresponding masses are
\begin{align}
    m_1^2 \equiv m_{\tilde{\psi}}^2 &= \frac{1}{4} \left(2m_\psi^2 + 2m_\chi^2 + y_H^2 v^2 - \sqrt{4(m_\chi^2 - m_\psi^2)^2 + y_H^2 v^2 (4m_\psi^2 + 4m_\chi^2 + y_H^2 v^2)} \right) \,, \\
    m_2^2 \equiv m_{\tilde{\chi}}^2 &= \frac{1}{4} \left( 2m_\psi^2 + 2m_\chi^2 + y_H^2 v^2 + \sqrt{4(m_\chi^2 - m_\psi^2)^2 + y_H^2 v^2 (4m_\psi^2 + 4m_\chi^2 + y_H^2 v^2)} \right) \,,
\end{align}
where we have defined $m_1 \equiv m_{\tilde{\psi}}$ and $m_2 \equiv m_{\tilde{\chi}}$. One can also show that $m_2 \geq m_1 + y_H v /\sqrt{2}$, with equality when $m_\chi = m_\psi$.  The mixing angles are given by
\begin{align}
    \tan(2\theta_L) &= 2\sqrt{2}\, y_H v \frac{m_\chi}{2(m_\chi^2 - m_\psi^2) - y_H^2 v^2} \,, \\
    \tan(2\theta_R) &= 2\sqrt{2}\, y_H v \frac{m_\psi}{2(m_\chi^2 - m_\psi^2) + y_H^2 v^2} \,.
\end{align}
The two mixing angles are related according to $m_1 \tan\theta_L = m_2 \tan\theta_R$ (such a relation is expected since the mass matrix contains only three real parameters:~$y_H v$, $m_\psi$ and $m_\chi$). The following useful relations also hold:
\begin{align}
    \cos\theta_L \sin\theta_R &= \frac{y_H v}{\sqrt{2}} \frac{m_1}{m_2^2 - m_1^2} \,, \\
    \sin\theta_L \cos\theta_R &= \frac{y_H v}{\sqrt{2}} \frac{m_2}{m_2^2 - m_1^2} \,.
\end{align}

The one-loop radiative masses, generated by the Feynman diagram in \cref{fig:feyn-class-2-after-ewsb} (left), are given by 
\begin{equation} \label{eq:mass2}
    m_{\tau,b} = \frac{y_\psi y_\chi}{16\pi^2} \frac{y_H v}{\sqrt{2}} n_2 n_3 F\left(x_1,x_2\right) \,,
\end{equation}
where now $x_{1,2} = m_{1,2}^2/m_\phi^2$, $F(x_1,x_2)$ is defined in \cref{eq:F}, and $n_2$, $n_3$ are model-dependent group theory factors. It is immediately clear that \cref{eq:mass2} is independent of the overall scale of new physics; however, as in Class~1, large masses lead to a fine-tuning of the Higgs mass, placing a theoretically motivated upper bound on the scale of new physics.  Since $F(x_1,x_2) \leq 1$, there is a lower bound on the product of Yukawa couplings needed to give the measured fermion masses. For the minimal $\tau$ model, with $n_2=n_3 = 1$, the bound is $y_\psi y_\chi y_H > 1.6$. 

The effective Yukawa coupling, \cref{fig:feyn-class-2-after-ewsb} (right), evaluated at $p_h^2=0$, is
\begin{equation} \label{eq:yeff2}
    y_{\tau,b}^\text{eff} = \frac{y_\psi y_\chi y_H}{16 \pi^2} n_2 n_3 \Bigg[ F(x_1,x_2) \\
    + \sin\theta_L \cos\theta_L \sin\theta_R \cos\theta_R \left( G(x_1, x_2) - \frac{m_1^2 + m_2^2}{m_1 m_2} F(x_1,x_2) \right) \Bigg] \,,
\end{equation}
with
\begin{multline}
    G(x_1,x_2) = \frac{1}{x_1 - x_2} \left( \frac{\left( x_2 (x_2 + 1) + x_1 (x_2 - 3) \right) x_2 \log x_2}{(x_2 - 1)^2} - \frac{\left( x_1 (x_1 + 1) + x_2 (x_1 - 3) \right) x_1 \log x_1}{(x_1 - 1)^2} \right) \\
    + 2\left( 2 + \frac{1}{x_1 - 1} + \frac{1}{x_2 - 1} \right) \,.
\end{multline}
The SM tree-level relation $m_f = y_fv/\sqrt{2}$ is violated by the radiatively generated masses and Yukawa couplings in \cref{eq:mass2,eq:yeff2}.  However, it is approximately recovered in the limit of small $\theta_L$, small $\theta_R$, or when $m_1\approx m_2$ ($m_\chi \approx m_\psi \gg v$).

%==============================================================================
\section{Class~2 - Benchmark models}
\label{sec:class2-pheno}
%==============================================================================

Our approach to exploring the phenomenology of this class of models follows the same strategy as for Class~1. While we explore three benchmark models in detail, many of the general features apply to all models of this class. The relevant experimental constraints are essentially the same as in Class~1, namely precision Higgs and electroweak measurements.  The modification of the Higgs coupling to the $\tau$-lepton/$b$-quark is given above, and the couplings to gauge bosons are discussed in \cref{app:higgs-couplings}.  The deviations in the $S$ and $T$ parameters are presented in \cref{app:STU} and the $Z$ couplings are discussed in \cref{app:Z-couplings}. 

Compared to Class~1, the most important difference in the Higgs couplings is that corrections to the $h\gamma\gamma$ and $hgg$ couplings are generally small in this class of models. This is due to the fact that the new contributions to the amplitudes decouple at least as fast as $1/M^3$, compared to $1/M$ in Class~1 models (where $M$ is the scale of new physics). Hence, the dominant effects in Higgs observables are due to the deviations of the $\tau$-lepton and $b$-quark effective Yukawa couplings from their SM values. Note, however, that in the $b$-quark models this can still lead to significant deviations across a range of Higgs measurements, due to the fact that $h\to\bar{b}{b}$ provides the dominant contribution to the Higgs total width.

Again there are theoretical constraints on the allowed regions of parameter space.  Since all of the new couplings are dimensionless, perturbative unitarity simply requires $y_H, y_\chi, y_\psi \lesssim \sqrt{4\pi}$.  Again, stronger bounds could be obtained by requiring the absence of Landau poles up to high scales.  If $\chi$ and $\psi$ are significantly above the weak scale they generate a fine-tuning problem for the Higgs.  We calculate this effect and define our fine-tuning measure in \cref{app:naturalness}.

The exotic particle number symmetry again allows for the possibility of a dark matter candidate, or leads to stringent bounds on the minimal models from searches for charged or coloured relics.  For each benchmark model we discuss these possibilities, and consider interactions which may break this stabilising symmetry.  We also briefly mention the current collider bounds relevant to these benchmark models.

%--------------------------------------------------------
\subsection{\texorpdfstring{Class~2 $\tau$-lepton Benchmark Model}{Class~2 tau-lepton Benchmark Model}}
%--------------------------------------------------------

We adopt the minimal model as our Class~2 $\tau$-lepton benchmark model. Specifically, we take
\begin{equation}
    \psi_{L,R} \sim (\mathbf{1},\mathbf{2},-\tfrac{3}{2})\,, \quad \chi_{L,R} \sim (\mathbf{1},\mathbf{1},-2) \,, \quad \phi \sim (\mathbf{1},\mathbf{1},-1) \,.
\end{equation}
The group theory factors for this model are $n_2 = 1$ and $n_3 = 1$.  With this choice of hypercharge assignment, there is clearly no dark matter candidate. There are, however, two terms that can break $U(1)_X$ and allow the exotic states to decay into the SM:~$\bar{e}_R^i H \psi_L$ and $\bar{L} L^c \phi$, which break $L$ and $X$ to $L+X$ and $L+3X$ respectively. In the former case $\bar{e}_R^i$ represents either an electron or muon; for the tau this term would break both $S_\chi$ and $S_\phi$. Such a coupling also induces mixing between $\psi$ and the charged lepton after electroweak symmetry breaking. 

%--------------------------------------------------------
\begin{figure}[ht]
    \centering
    \includegraphics[height=0.45\textwidth]{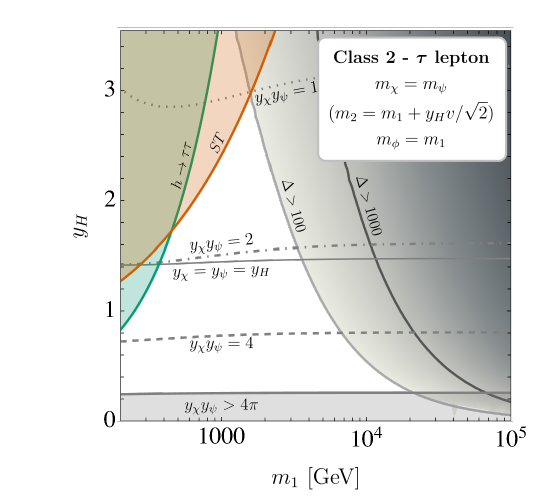}
    \includegraphics[height=0.45\textwidth]{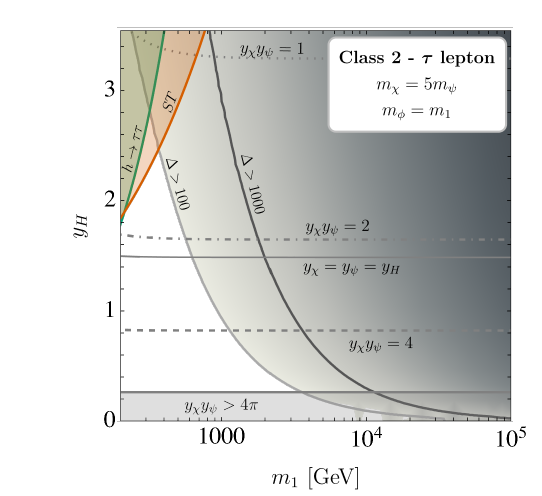}
	\caption{Theoretical (grey) and current $2\sigma$ experimental constraints (coloured) on the Class~2 $\tau$-lepton benchmark model.}
	\label{fig:2tau}
\end{figure}
%--------------------------------------------------------

In \cref{fig:2tau} we show the constraints on this model for two slices of parameter space.  In the left panel we fix the Lagrangian mass parameters to be equal, $m_\chi = m_\psi$, which corresponds to the smallest possible value for $m_2$ for a given $m_1$ and $y_H$; in the right panel we take $m_\chi = 5 m_\psi$.  In both panels the scalar mass is fixed to be $m_\phi = m_1$. 

The theoretical constraints are shown in grey.  The product of Yukawa couplings $y_\chi y_\psi$ which reproduces the tau mass is shown at $y_\chi y_\psi = $ 1, 2, 4, $4\pi$, and where $y_\chi = y_\psi = y_H$.  In the left panel we see that the tau mass can indeed be radiatively generated at one-loop with order one Yukawas ($y_\chi y_\psi y_H \approx 3$). This is slightly above the absolute bound $y_\chi y_\psi y_H > 1.6$ given below \cref{eq:mass2}, which simply required $F(x_1,x_2)\leq 1$.  Fine-tuning generally becomes a concern at lower mass scales than in the Class~1 $\tau$-lepton benchmark model.  Taking all Yukawas to be equal (around 1.4), a fine-tuning of 1\% is required if the new physics scale is around 3\,TeV, while a fine-tuning of 0.1\% is required for new physics at 10\,TeV.  In the right-hand slice of parameter space, the Yukawa couplings have to be slightly larger, and significant fine-tuning is required at even lower masses.

The current experimental $2\sigma$ constraints are shown as coloured regions.  In the left slice, the deviation in the Higgs to $\tau^+\tau^-$ signal strength is most constraining at low masses, while the $S$ and $T$ parameters take over above 400\,GeV.  In the right panel, both constraints weaken due to the larger value of $m_2$, and the $S$ and $T$ parameters provide the best constraint everywhere.  For the reasons discussed above, there is no constraint from $h \to \gamma\gamma$ or $h \to ZZ$, and measurements of the $Z\bar{\tau}\tau$ couplings are not currently sensitive enough to provide a constraint.

%--------------------------------------------------------
\begin{figure}[ht]
    \centering
    \includegraphics[height=0.45\textwidth]{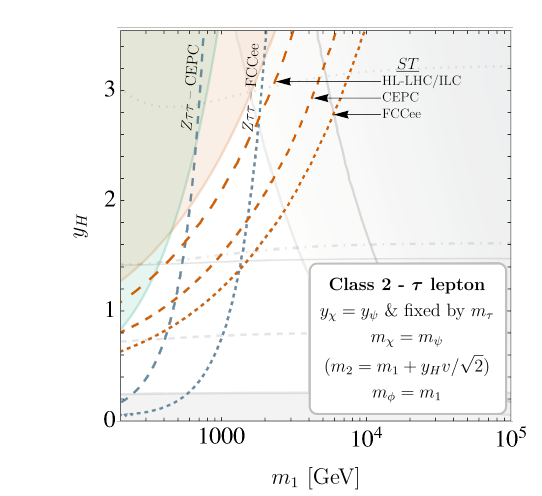}
    \includegraphics[height=0.45\textwidth]{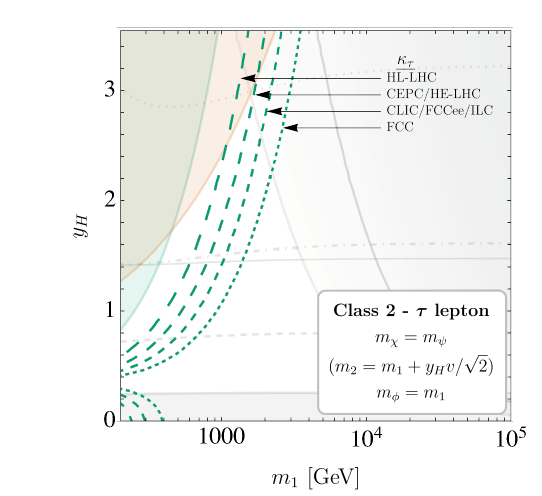}
	\caption{Projected $2\sigma$ experimental sensitivities on the Class~2 $\tau$-lepton benchmark model from future colliders, for the left slice of parameter space in \cref{fig:2tau}.}
	\label{fig:2tau-FC}
\end{figure}
%--------------------------------------------------------

In \cref{fig:2tau-FC} we show the parameter space that can be probed at future colliders, for the slice of parameter space shown in the left panel of \cref{fig:2tau}.  Future improvements in the measurements of $S$ and $T$ can probe significant regions of parameter space, with the FCCee reaching into the multi-TeV range.  Measurements of the $Z\bar{\tau}\tau$ couplings provide a complementary probe of the parameter space, with the FCCee able to probe masses above 1\,TeV.  Here we have taken $y_\chi = y_\phi$; variations from this typically increase the future sensitivity.  We see that, ultimately, electroweak observables will have the best sensitivity to this model; however, in the shorter term measurements of the $h\bar{\tau}\tau$ coupling at the HL-LHC (and potentially the ILC) can probe new regions of parameters space. Furthermore, if deviations were seen in electroweak observables, Higgs coupling measurements would be essential to establish the connection to fermion mass generation.

Direct collider searches for the exotic particles are sensitive to the details of the $U(1)_X$ breaking, but one generally expects relatively weak limits in this model due to the absence of coloured states. The exception is if the $U(1)_X$-breaking couplings are sufficiently small that the lightest exotic is collider-stable. There are then strong bounds from LHC searches for heavy stable charged particles produced via Drell-Yan. For a fermion with charge two, ATLAS~\cite{1812.03673} excludes the mass range $50-980$\,GeV, while for a singly charged fermion CMS~\cite{1609.08382} excludes $50-650$\,GeV.

%--------------------------------------------------------
\subsection{\texorpdfstring{Class~2 $b$-quark Benchmark Model}{Class~2 b-quark Benchmark Model}}
%--------------------------------------------------------

We also adopt the minimal model as our $b$-quark benchmark. The fermions carry the same $SU(3)_C$ and $SU(2)_L$ quantum numbers as in the $\tau$ case, but now the scalar is charged under $SU(3)_C$,
\begin{equation}
    \psi_{L,R} \sim (\mathbf{1},\mathbf{2},\tfrac{1}{2})\,, \quad \chi_{L,R} \sim (\mathbf{1},\mathbf{1},0) \,, \quad \phi \sim (\mathbf{3^*},\mathbf{1},\tfrac{1}{3}) \,.
\end{equation}
The group theory factors for this model are $n_2 = 1$ and $n_3 = 1$.  We have chosen the hypercharge assignments such that the lighter $\chi-\psi$ eigenstate is neutral and a dark matter candidate. It can be either a Dirac or Majorana fermion, with a Majorana mass for $\chi$ breaking $U(1)_X \to \mathcal{Z}_2$. Although it doesn't strongly impact the collider observables, for concreteness we assume it to be Dirac. Alternatively, there are several possibilities for $U(1)_X$-breaking couplings with SM particles in this model, which can open up regions of parameter space where there is no viable dark matter candidate. These are $\overline{\psi_R^c}He_R^i$, $\bar{\chi}_RHL_L$, $\overline{\chi_L^c}HL_L$, $\overline{(e_R^i)^c}u_R^j\phi$, and $\overline{(d_R^i)^c}(u_R^j)^c\phi$; in the final term $d_R^i$ corresponds to either $d_R$ or $s_R$, since for $b_R$ this term would break both the $S_\chi$ and $S_\phi$ symmetries. Each of these terms also breaks either baryon or lepton number, but not both. 

%--------------------------------------------------------
\begin{figure}[ht]
    \centering
    \includegraphics[height=0.45\textwidth]{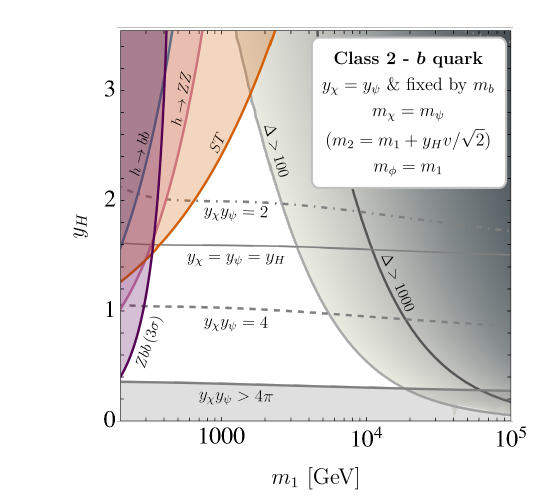}
    \includegraphics[height=0.45\textwidth]{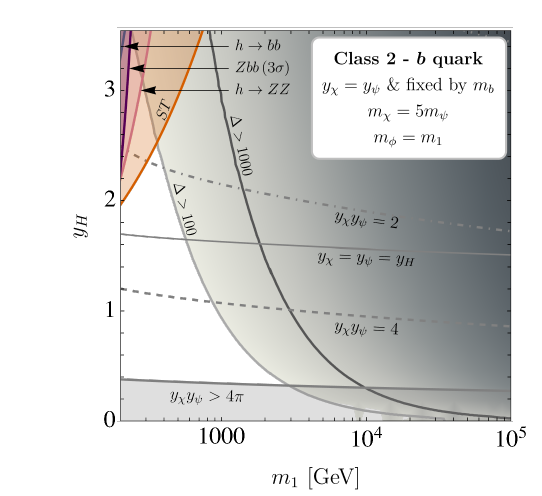}
	\caption{Theoretical (grey) and current $2\sigma$ experimental constraints (coloured) on the Class~2 $b$-quark benchmark model.  The $Z\bar{b}b$ constraint is given at $3\sigma$ since there is a $\gtrsim2\sigma$ tension in the current electroweak fit.}
	\label{fig:2b}
\end{figure}
%--------------------------------------------------------

In \cref{fig:2b} we show the current constraints on two slices of parameter space.  The value of $y_\chi y_\psi$ needed to give the $b$-quark mass is slightly larger than was required for the $\tau$ benchmark model above since the $b$-quark mass is slightly larger, even after taking into account the renormalisation group running described previously.  The Yukawas can simultaneously be around 1.6.  Again, generation of the $b$-quark mass at one-loop with order one Yukawas is appropriate.  The degree of fine-tuning is identical to the Class~2 $\tau$ benchmark model, as the exotic fermions, which provide the mass correction to the Higgs, are identical up to hypercharge.

While the constraint from $S$ and $T$ is almost identical to the $\tau$ model, with a tiny difference due to the different hypercharges, the other experimental constraints are quite different.  The Higgs to $\bar{b}b$ signal strength constraint is somewhat weaker than the analogous constraint in the $\tau$ case. This is partly due to the differing experimental central values (the uncertainties are comparable) but also due to the partial cancellation of the $h\bar{b}b$ coupling deviation in the branching ratio, as discussed for Class~1.  There are now non-negligible constraints from Higgs to diboson, with Higgs to $ZZ$ giving the strongest constraint among these.  Although this model does not alter the production or decay of the Higgs in this channel, it can significantly alter the total width of the Higgs which leads to a deviation in the signal strength.  Finally, we see that there is a constraint from measurements of the $Z\bar{b}b$ couplings.  We show the constraint at $3\sigma$ since there is a $\gtrsim2\sigma$ tension in the electroweak fit, and the new physics contributions increase the tension.  As above, we set $y_\chi = y_\psi$ in plotting this constraint; deviations from this can both strengthen and weaken the constraint.

In the right-hand panel of \cref{fig:2b} we see that increasing the mass of $m_2$ weakens the experimental constraints but slightly strengthens the perturbativity constraint on $y_\chi y_\psi$ and increases the degree of fine-tuning at large masses.

%--------------------------------------------------------
\begin{figure}[ht]
    \centering
    \includegraphics[height=0.45\textwidth]{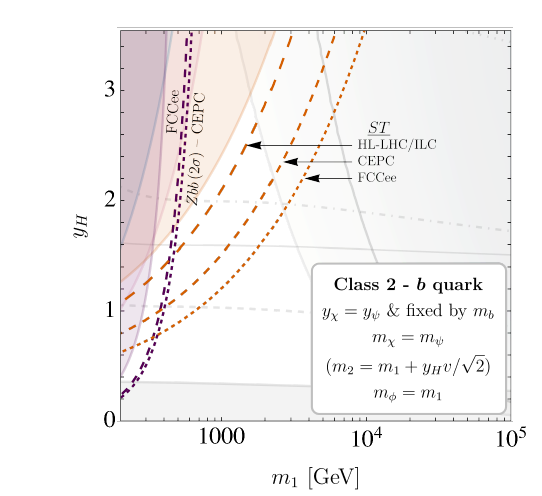}
    \includegraphics[height=0.45\textwidth]{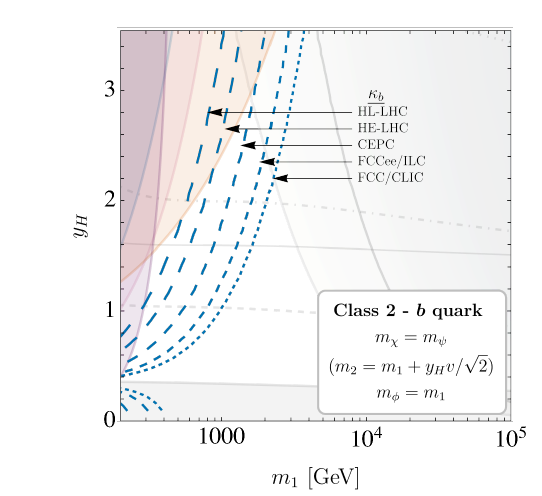}
	\caption{Projected $2\sigma$ experimental sensitivities on the Class~2 $b$-quark benchmark model from future colliders, for the left slice of parameter space in \cref{fig:2b}.}
	\label{fig:2b-FC}
\end{figure}
%--------------------------------------------------------

In \cref{fig:2b-FC} we show the projected sensitivity of future colliders.  In the left panel we show the improvement from electroweak observables.  While there is not a substantial gain from the $Z\bar{b}b$ couplings, the $S$ and $T$ parameters can cut into significant regions of parameter space.  Improvements in Higgs measurements, shown in the right panel, will also provide a powerful probe and can reach higher fermion masses than the electroweak measurements at low $y_H$.

We now comment on the dark matter phenomenology. Assuming $m_\chi<m_\psi$ and neglecting $\chi-\psi$ mixing, the relic abundance is set via annihilation into $\bar{b}b$ via $t$-channel $\phi$ exchange. The thermally-averaged cross-section is then
\begin{equation}
    \langle \sigma v \rangle = \frac{1}{32\pi m_\chi^2} \frac{y_\chi^4}{(1 + m_\phi^2/m_\chi^2)^2} + \mathcal{O}(v^2) \,.
\end{equation}
Given the large values of $y_\chi$ needed to obtain the $b$-quark mass, the observed dark matter abundance should be attainable with $\chi$ masses up to a few TeV, depending on the mass of $\phi$.  For higher masses the relic abundance would be too large and would overclose the universe.  Direct and indirect detection constraints for this model warrant a dedicated analysis. 

Turning briefly to direct collider searches, the coloured scalar $\phi$ is constrained by production at the LHC. Assuming $U(1)_X$ is unbroken, one can re-purpose an ATLAS sbottom search~\cite{1708.09266} to obtain the bound $m_\phi>950$\,GeV, provided $m_1<420\,$GeV. On the other hand, if the stabilising symmetry is broken then $\phi$ can decay into a wide range of final states, depending on the $U(1)_X$-breaking term.

%--------------------------------------------------------
\subsection{\texorpdfstring{Class~2 Combined $b+\tau$ Benchmark Model}{Class~2 Combined b + tau Benchmark Model}}
%--------------------------------------------------------

Finally, we consider a Class~2 benchmark model which can generate both the $\tau$ mass and the $b$ mass simultaneously.  We consider the minimal combined model with the field content
\begin{equation}
    \psi_{L,R} \sim (\mathbf{3^*},\mathbf{2},-\tfrac{1}{6})\,, \quad \chi_{L,R} \sim (\mathbf{3^*},\mathbf{1},-\tfrac{2}{3}) \,, \quad \phi \sim (\mathbf{3},\mathbf{1},-\tfrac{1}{3}) \,.
\end{equation}
The group theory factors for this model are $n_2^\tau = 1$, $n_3^\tau = 3$, $n_2^b = 1$ and $n_3^b = 2$.  As discussed in \cref{sec:class2-classification-combined}, the hypercharges are fixed in the combined models.  Since all fields are coloured and charged, there is no obvious dark matter candidate. There can be additional Lagrangian terms containing SM fields which break the $\mathcal{Z}_2$ stabilising symmetry. While these terms also break $B$ and $L$, leading to proton decay, the experimental bounds can be satisfied as we discuss below.

%--------------------------------------------------------
\begin{figure}[ht]
    \centering
    \includegraphics[height=0.45\textwidth]{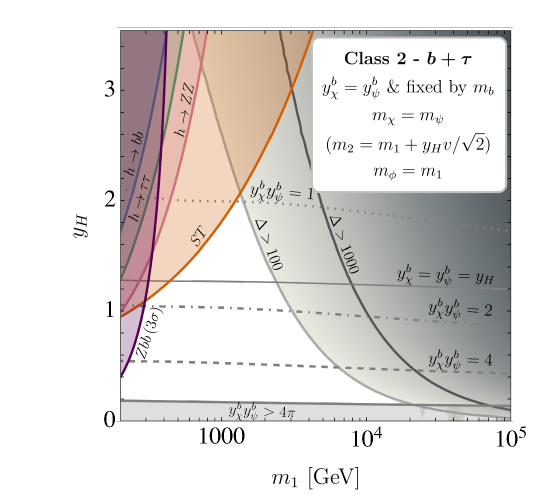}
    \includegraphics[height=0.45\textwidth]{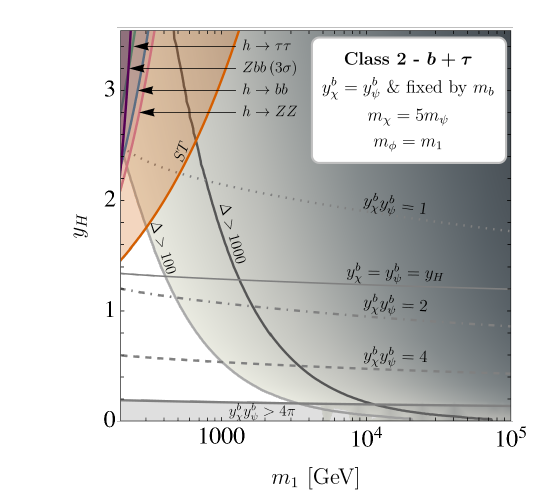}
	\caption{Theoretical (grey) and current $2\sigma$ experimental constraints (coloured) on the Class~2 combined $b+\tau$ benchmark model.}
	\label{fig:2comb}
\end{figure}
%--------------------------------------------------------

In \cref{fig:2comb} we show the theoretical and existing $2\sigma$ constraints on the same two slices of parameter space.  The value of $y_\chi^b y_\psi^b$ required to reproduce the running $b$-quark mass is larger than the required tau couplings $y_\chi^\tau y_\psi^\tau$ due to $m_b > m_\tau$ and $n_3^b < n_3^\tau$, which both act to increase $y_\chi^b y_\psi^b$ in \cref{eq:mass2}.  The $b$-quark mass can be obtained with $y_\chi^b = y_\psi^b = y_H \approx 1.3$ and the $\tau$-lepton mass can be obtained with $y_\chi^\tau = y_\psi^\tau = y_H \approx 1.0$.  We see that the required Yukawa couplings are smaller than in the Class~2 $b$-quark benchmark model, due to the larger group theory factor.  The fine-tuning measure is larger here than in the Class~2 $\tau$-lepton and $b$-quark models due to the larger $SU(3)_C$ representation of the exotic fermions, although it makes a negligible difference in the plot.

While this model has a wide range of existing experimental constraints from precision Higgs measurements, these are weaker than those from the $S$ and $T$ parameters and the deviations in the $Z\bar{b}b$ couplings.  The $S$ and $T$ parameters provide a stronger constraint that in the Class~2 $\tau$-lepton and $b$-quark benchmark models due to the larger $SU(3)_C$ representation of the exotic fermions.  The $Z\bar{b}b$ constraint is identical to that in the Class~2 $b$-quark model.  Note again that this constraint depends on $y_\chi^b$ and $y_\phi^b$ independently, and not only on the product.  As above we set $y_\chi^b = y_\phi^b$ in plotting the constraint.  In the right panel we see again that the effect of increasing $m_2$ is to weaken the experimental constraints, but increase the required Yukawa couplings and the degree of fine-tuning.  

%--------------------------------------------------------
\begin{figure}[ht]
    \centering
    \includegraphics[height=0.45\textwidth]{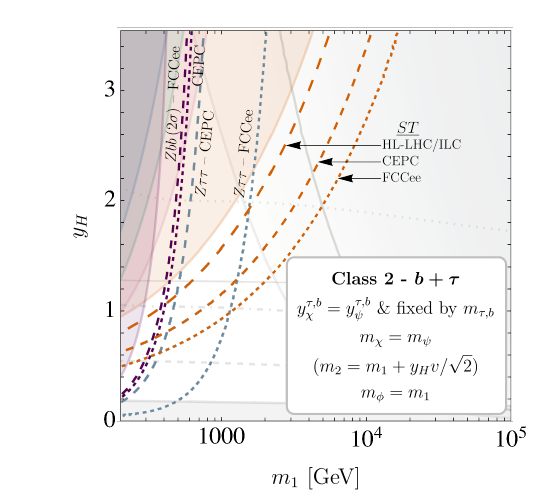}
    \includegraphics[height=0.45\textwidth]{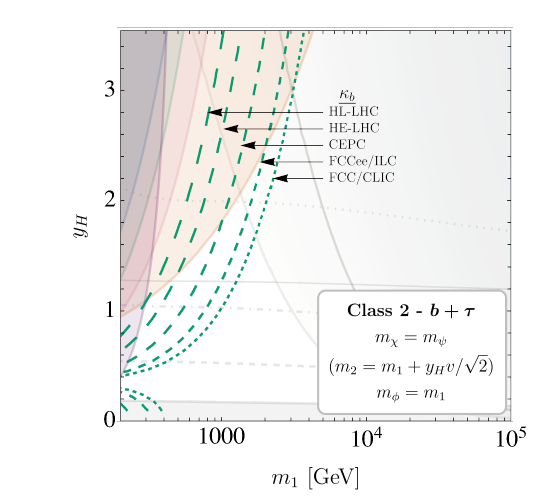}
    \\
    \includegraphics[height=0.45\textwidth]{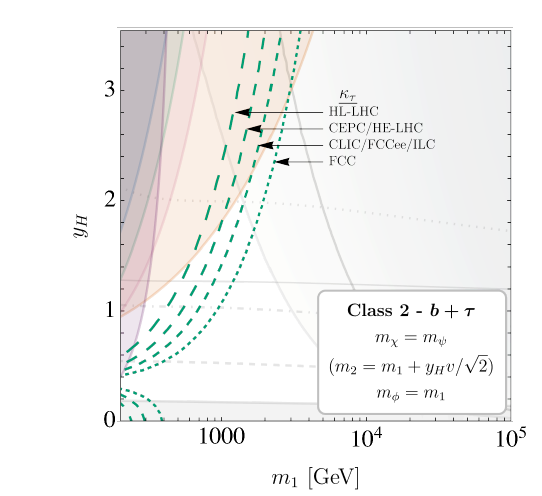}
	\caption{Projected $2\sigma$ experimental sensitivities on the Class~2 combined $b+\tau$ benchmark model from future colliders, for the left slice of parameter space in \cref{fig:2comb}.  In the top-left panel we have assumed $y_\chi^{\tau,b} = y_\psi^{\tau,b}$ and fixed the values to reproduce the fermion masses.}
	\label{fig:2comb-FC}
\end{figure}
%--------------------------------------------------------

The sensitivities of future colliders are shown in \cref{fig:2comb-FC}.  The electroweak $S$ and $T$ parameters will continue to provide the best probe of the parameter space, with the $Z\bar{\tau}\tau$ coupling measurements playing a complementary role at small $y_H$.
The precision Higgs measurements, while not as far reaching as the electroweak measurements, would provide important evidence of the connection to fermion mass generation if a departure from the SM were seen.

To avoid the bounds on cosmological relics, the $\mathcal{Z}_2$ symmetry should be broken. This can be achieved via the terms $\overline{\psi_R^c}Hd_R^i$, $\overline{\chi_L^c}HQ_L$, $\overline{e}_R^i(u_R^j)^c\phi$, $\overline{(d_R^i)^c}u_R^j\phi$, $\overline{\psi_L^c}Q^i$, or $\overline{\chi_R^c}u_R$. All of these terms break the generalised $B$ and $L$ symmetries to the combination $B-L$, and lead to proton decay. For either of the last two terms above this occurs via the lowest possible dimension operator (dimension-6, as $B-L$ is preserved), mediated at tree-level by $\phi$. Consider the term\footnote{This softly breaks $S_\psi$ and $S_\chi$, but any effects on fermion mass generation are negligible.} $\mu_\slashed{\mathcal{Z}_2} \overline{\chi_R^c}b_R$ (the discussion proceeds similarly for $\overline{\psi_L^c}Q_L^3$). As in Class~1, we compute the third family process as a conservative estimate of the proton decay rate, and obtain
\begin{equation}
    \Gamma_p \sim \frac{|y_\chi^b y_\chi^\tau|^2}{512\pi^3} \left(\frac{\mu_\slashed{\mathcal{Z}_2}}{m_\chi}\right)^4 \frac{m_p^5}{m_\phi^4} \,.
\end{equation}
Requiring that the exotics decay before BBN leads to a lower bound on $\mu_\slashed{\mathcal{Z}_2}$:
\begin{equation}
    \left(\frac{\mu_\slashed{\mathcal{Z}_2}}{m_\chi}\right)^2 \gtrsim 10^{-25} \left(\frac{\text{TeV}}{M} \right) \,,
\end{equation}
where $M$ is the mass of the decaying exotic and we have used $y_\chi^b \sim y_\psi^\tau\sim g_s \sim 1$. Saturating this bound, and setting all exotic masses equal to $M$, leads to a proton lifetime of
\begin{equation} \label{eq:class2-proton-lifetime}
  \tau_p \sim 10^{35}  \left(\frac{M}{\text{TeV}}\right)^6 \text{yr} \,.
\end{equation}
The mass-dependence is milder than in Class~1, due to the fact that here proton decay proceeds via a dimension-6, rather than a dimension-7, operator. In any case, the experimental bounds discussed in \cref{sec:class1-pheno-combined} are satisfied provided the new physics is above the weak scale; larger values of $\mu_\slashed{\mathcal{Z}_2}$ could lead to observable proton decay. Alternatively, the bounds can be avoided by extending the minimal model, for example by including a dark matter candidate, which avoids introducing the small scale $\mu_\slashed{\mathcal{Z}_2}$.

Given the proton decay bounds on the $\mathcal{Z}_2$-breaking couplings, one expects the lightest exotic to be collider-stable. As in Class~1 combined models, the relevant searches are for $R$-hadrons~\cite{1902.01636}. If $\phi$ is the lightest exotic then the bound is $m_\phi>1250$\,GeV~\cite{1902.01636} from sbottom searches. Recasting should give similar bounds if $\psi$ or $\chi$ are the lightest exotic.

%==============================================================================
\section{Conclusions}
%==============================================================================

In this study we set out to determine whether the minimal fermion mass generation mechanism of the SM has been experimentally established for the $\tau$-lepton or the $b$-quark by recent measurements of the Higgs couplings to fermions, or whether it will be established in the foreseeable future.  By classifying minimal one-loop radiative models and considering representative benchmark models, we have found that the answer is no, on both counts.  Whether this broad conclusion is positive or negative is perhaps a matter of taste.  While we will not be able to rule out one-loop radiative generation of the $\tau$ or $b$ mass without new ways to probe these models, there remains the possibility that these models may be part of the answer to the flavour problem and furthermore that they may hold answers for other fundamental questions such as the particle nature of dark matter.  

Our systematic classification, the first of its kind for these models, targets the minimal one-loop radiative models containing new fermions and scalars. These models divide into two distinct classes based on the field content and the topology that generates the effective Yukawa couplings.  We analysed the symmetries satisfied by the required Lagrangian terms and determined all possible quantum numbers of the exotic particles.  This allowed us to list the minimal models, which contain only singlets and fundamentals of the non-Abelian SM gauge groups.  We also highlight particularly economical models which can simultaneously generate both the $\tau$-lepton mass and the $b$-quark mass with the same new degrees of freedom.

The exotic particles in these models can in principle be arbitrarily massive; however, this comes at the expense of fine-tuning of the weak scale. While measures of such fine-tuning are inherently subjective, mass scales close to the weak scale are theoretically favoured and have the potential to be tested in the future. Current observations allow for perturbative and non-fine-tuned models that can radiatively generate the $\tau$-lepton mass, or the $b$-quark mass, or both simultaneously. While a non-discovery at future colliders would push the required Yukawa couplings to be larger than 1 in some minimal models, they could still be smaller than $\approx 1.5$. 

A key prediction of radiative models is that they violate the SM relation between a fermion's mass and its coupling to the physical Higgs boson.  This relation is currently being tested at the LHC, and the HL-LHC and future Higgs factories will provide an important test of these models in the coming years.  Class~1 models also predict observable deviations in Higgs couplings to gluons and photons. The electroweak $S$ and $T$ parameters provide strong constraints, and deviations in the $Z$ boson couplings to $\tau$-leptons and $b$-quarks are also predicted. If only the minimal couplings required for fermion mass generation are present, then the lightest exotic state is stable. While additional interactions with SM fields may break the stabilising symmetry, this highlights the intriguing possibility that dark matter could be intricately connected with fermion mass generation.

We studied the phenomenology of six benchmark models, which radiatively generate the $\tau$-lepton mass, the $b$-quark mass, or both, for the Class~1 and Class~2 topologies.  For the Class~1 $\tau$-lepton benchmark, we found that the strongest experimental constraints are currently from the Higgs to $\tau^+\tau^-$ signal strength and the electroweak $S$ and $T$ parameters.  In the future, measurements of the $Z\bar{\tau}\tau$ couplings will also be competitive.  This model can contain a viable dark matter candidate and current collider bounds for the exotic particles are expected to be at the several hundred GeV level.

For the Class~1 $b$-quark benchmark we considered a model with $SU(3)_C$ triplets.  In this model measurements of the Higgs to $ZZ$ signal strength are most constraining.  Future collider measurements can probe an interesting region of parameter space, but regions with Yukawas = 1 will still be allowed.  This model does not contain a dark matter candidate, and additional couplings to SM particles can lead to existing collider limits in the 1--2\,TeV range.

The Class~1 combined $\tau + b$ benchmark model contains the same number of new degrees of freedom as the Class~1 $b$-quark benchmark model.  In this model the Higgs couplings to $\bar{\tau}\tau$, $\bar{b}b$, $gg$ and $\gamma\gamma$ all deviate from their SM expectations, and the strongest constraint comes from the Higgs to diphoton signal strength.  However, even with improvements at future colliders there will still be unexplored parameter space with Yukawas = 1.  While there is no dark matter candidate in this model, there are several interactions with SM particles that can break the stabilising symmetry, but these all lead to proton decay.  While it is possible to satisfy existing limits and ensure that the exotic particles decay before Big Bang Nucleosynthesis, direct collider searches give bounds in the 1--2\,TeV range for the long-lived exotic.

The Class~2 models have a different topology, with the exotic fermions coupling to the SM Higgs.  The minimal Class~2 $\tau$ benchmark model has viable regions of parameter space with all Yukawa couplings $\approx$ 1.4 and with the exotic particle masses below a TeV.  This parameter space will be probed to around 2\,TeV by precision measurements at future colliders, while greater than 1\% fine-tuning is required above around 3\,TeV.  Although there is no dark matter candidate, the stabilising symmetry of the exotics can be broken by couplings to SM particles, which may lead to bounds from direct searches. 

In studying the minimal Class~2 $b$-quark benchmark model, we saw that the Yukawa couplings had to be slightly larger, due to the larger $b$ mass.  In the region where all Yukawas are equal, future colliders will again probe masses up to $\sim2$\,TeV, while greater than 1\% fine-tuning is required above $\sim3$\,TeV.  This model either contains a dark matter candidate or can accommodate couplings to SM particles which break the stabilising symmetry, removing relic abundance constraints but creating new collider probes.

Finally, we analysed the minimal Class~2 combined $\tau + b$ benchmark model.  In this model all exotic fields are $SU(3)_C$ triplets or anti-triplets, leading to smaller Yukawa couplings than seen in the other Class~2 benchmark models.  The model has viable parameter space with all Yukawas $\approx$ 1.3, and the most significant current constraints are electroweak measurements, both the $S$ and $T$ parameters and the $Z$ couplings to $b$-quarks.  Future colliders will probe this model into the multi-TeV scale. Unless a dark matter candidate is added to the model, there are also stringent bounds from either the cosmological relic or proton decay.  If the $\mathcal{Z}_2$ symmetry is broken, the proton decay bounds imply that one of the exotics is long-lived and subject to collider bounds at the TeV scale.

%==============================================================================
\section{Acknowledgements}
%==============================================================================

The authors would like to thank John Gargalionis for useful discussion and collaboration in the initial stages of this project.  This work was supported by the Australian Government through the Australian Research Council.

%==============================================================================
\appendix
%==============================================================================

%==============================================================================
\section{Higgs Couplings}
\label{app:higgs-couplings}
%==============================================================================

Class~1 and Class~2 models generically give modified predictions for Higgs production and decay. These effects are described by the effective Yukawa couplings in \cref{eq:yeff1,eq:yeff2}, along with new one-loop contributions to the effective couplings to gauge bosons. The latter can be neglected for the $W$ and $Z$, since the corrections to the SM tree-level couplings are negligible, but are important for gluons and photons. It's convenient to define the following effective Lagrangian
\begin{equation}
    \mathcal{L}_\text{eff} = -\frac{c_\gamma \alpha_0 }{4\pi} h F^{\mu\nu}F_{\mu\nu} - \frac{c_g \alpha_s}{4\pi} h G_a^{\mu\nu}G^a_{\mu\nu} \,,
\end{equation}
where $c_\gamma$, $c_g$ are momentum-dependent effective couplings, and $\alpha_0$ is the electromagnetic fine-structure constant in the Thomson limit\footnote{This avoids large logarithms for on-shell photons in the higher-order corrections to the $h\to\gamma\gamma$ partial width.}.

In Class~1 models, one-loop scalar diagrams give (for $p_h^2=m_h^2$)
\begin{align}
    c_\gamma^\text{NP} &= d_3(\phi) \bigg(\sum_{i=1}^{n_2} Q_i^2\bigg)  \frac{a \sin(2\theta)}{4\sqrt{2} m_h^2} \left( A_0\left(\frac{4m_2^2}{m_h^2}\right) - A_0\left(\frac{4m_1^2}{m_h^2}\right) \right) \,, \\
    c_g^\text{NP} &= d_2(\phi) T_3(\phi) \frac{a \sin(2\theta)}{4\sqrt{2} m_h^2} \left( A_0\left(\frac{4m_2^2}{m_h^2}\right) - A_0\left(\frac{4m_1^2}{m_h^2}\right) \right) \,,
\end{align}
where the sum is over the electric charges, $Q_i$, of the mixed $\phi-\eta$ states and $T_3(\phi)$ is the index of the $SU(3)_C$ representation of $\phi$, defined by $\text{Tr}(t_R^a t_R^b)=T(R) \delta^{ab}$. Notice that the amplitude has an overall enhancement of $a/m_h$ relative to the SM contribution; however, the amplitude still decouples as the overall mass scale of new physics increases, due to the cancellation between the two $A_0$ terms.

In Class~2, the one-loop fermion contributions are
\begin{align}
    c_\gamma^\text{NP} &= d_3(\psi) \bigg(\sum_{i=1}^{n_2} Q_i^2\bigg)  \frac{y_H}{8\sqrt{2}} \bigg( \frac{\sin\theta_L \cos\theta_R}{m_2} A_{1/2}\left(\frac{4m_2^2}{m_h^2}\right) - \frac{\cos\theta_L \sin\theta_R}{m_1} A_{1/2}\left(\frac{4m_1^2}{m_h^2}\right) \bigg) \,, \\
    c_g^\text{NP} &= n_2 T_3(\psi)\frac{y_H}{8\sqrt{2}} \bigg( \frac{\sin\theta_L \cos\theta_R}{m_2} A_{1/2}\left(\frac{4m_2^2}{m_h^2}\right) - \frac{\cos\theta_L \sin\theta_R}{m_1} A_{1/2}\left(\frac{4m_1^2}{m_h^2}\right) \bigg) \,,
\end{align}
where here the sum is over the electric charges of the mixed $\psi-\chi$ states. These contributions decouple much faster than in Class~1 ($1/M^3$ vs $1/M$ with mixing angles held fixed), and so are generally negligible. 

The standard loop functions are 
\begin{gather}
    A_0(x) = -x f(x) \,, \\
    A_{1/2}(x) = -2x (1 + (1-x) f(x)) \,, \\
    \intertext{with}
    f(x) = 
    \begin{cases} 
      -\frac{1}{4} \left(\log \frac{1+\sqrt{1-x}}{1-\sqrt{1-x}} -i\pi \right)^2 & x < 1 \,, \\
      \arcsin^2{\frac{1}{\sqrt{x}}} & x \geq 1 \,.
   \end{cases}
\end{gather}

The ATLAS and CMS experiments have measured the signal strengths ($\sigma\times BR$ normalised to the SM) for a variety of Higgs production and decay modes. The most constraining observables in these models are
\begin{align}
    \mu_{bb}^{VH} &\simeq 
    \frac{|y_b^\text{eff}|^2}{|y_b^\text{SM}|^2} \frac{1}{\kappa_h^2} 
    = \frac{\kappa_b^2}{\kappa_h^2}
    \,, \\
    \mu_{\tau\tau}^{ggH} &\simeq \frac{|c_g^\text{SM} + c_g^\text{NP}|^2}{|c_g^\text{SM}|^2}  \frac{|y_\tau^\text{eff}|^2}{|y_\tau^\text{SM}|^2} \frac{1}{\kappa_h^2}  = \frac{\kappa_g^2\kappa_\tau^2}{\kappa_h^2} \,, \\
    \mu_{\gamma\gamma}^{ggH} &\simeq \frac{|c_g^\text{SM} + c_g^\text{NP}|^2}{|c_g^\text{SM}|^2} \frac{|c_\gamma^\text{SM} + c_\gamma^\text{NP}|^2}{|c_\gamma^\text{SM}|^2} \frac{1}{\kappa_h^2}  = \frac{\kappa_g^2\kappa_\gamma^2}{\kappa_h^2} \,, \\
    \mu_{ZZ}^{ggH} &\simeq \frac{|c_g^\text{SM} + c_g^\text{NP}|^2}{|c_g^\text{SM}|^2} \frac{1}{\kappa_h^2}  = \frac{\kappa_g^2}{\kappa_h^2} \,,
\end{align}
where we have defined the $\kappa_i$ as the usual ratios of couplings and
\begin{align}
    \kappa_h^2 \equiv \frac{\Gamma_h}{\Gamma^\text{SM}_h}&=
    \frac{1}{\Gamma^\text{SM}_h}\left(\kappa_b^2\, \Gamma^\text{SM}_{h\to bb} 
    + \kappa_g^2\, \Gamma^\text{SM}_{h\to gg} 
    +  \kappa_\tau^2\, \Gamma^\text{SM}_{h\to \tau\tau} 
    +  \kappa_\gamma^2\, \Gamma^\text{SM}_{h\to \gamma\gamma} 
    + \Gamma^\text{SM}_{h\to\text{other}}\right)\,.\label{eq:mu-h}
\end{align}
For the SM contributions to $c_g$ we include QCD corrections up to NNLO~\cite{hep-ph/9708255}, while for $c_\gamma$ we include the NLO QCD corrections in the large top mass limit~\cite{hep-ph/9504378}. Note that the soft QCD corrections to the gluon-fusion production cross-section cancel in the above ratios. The current best-fit values are
\begin{align}
    \mu_{\tau\tau}^{ggH} &= 1.15^{+0.16}_{-0.15} \text{~\cite{Zyla:2020zbs}} \,, \notag \\
    \mu_{bb}^{VH} &= 1.04 \pm 0.13 \text{~\cite{Zyla:2020zbs}} \,, \notag \\
    \mu_{\gamma\gamma}^{ggH} &= 1.00^{+0.09}_{-0.07} \text{~\cite{ATLAS:2020pvn,CMS:2020omd}} \,, \notag \\
    \mu_{ZZ}^{ggH} &= 0.97^{+0.09}_{-0.08} \text{~\cite{2004.03447,CMS:2019chr}} \,,
\end{align}
where for $\mu_{\gamma\gamma}^{ggH}$ and $\mu_{ZZ}^{ggH}$ we have performed a naive combination of the most recent experimental measurements.

%==============================================================================
\section{\texorpdfstring{$S$, $T$, $U$ Parameters}{S, T, U Parameters}}
\label{app:STU}
%==============================================================================

Constraints on new physics from electroweak precision data can be characterised using the 
Peskin-Takeuchi parameters~\cite{Peskin:1991sw}, provided that the new physics is above the weak scale and the dominant effects occur via vacuum polarisation. Although the models in Classes~1~\&~2 also give vertex corrections at one-loop, these affect only the third-family fermions and furthermore are found to be numerically small, justifying the use of the Peskin-Takeuchi parameters. We use the results from the electroweak fit in Ref.~\cite{1608.01509} (with $U=0$) which constrains new physics contributions to satisfy
\begin{equation}
    S = 0.10 \pm 0.08 \,, \quad T = 0.12 \pm 0.07 \,, \\
\end{equation}
at $1\sigma$ with a correlation of 0.86.

In the limit $m_{NP} \gg m_Z$, the $S$ and $T$ parameters are directly related to the Wilson coefficients of the dimension-six operators $\mathcal{O}_S=H^\dagger \sigma^i H W_{\mu\nu}^i B^{\mu\nu}$ and $\mathcal{O}_T=|H^\dagger D_\mu H|^2$ ($U$ corresponds to a dimension-eight operator and is negligible in this limit). Below, we provide the expressions for $S$ and $T$ in this limit, where they simplify significantly; however, our numerical results in \cref{sec:class1-pheno,sec:class2-pheno} use the complete one-loop expressions.

In Class~1 models, only the scalars couple directly to the Higgs and give a non-zero contribution to $S$ and $T$ at one-loop. Assuming $d_2(\phi) > d_2(\eta)$ (the alternate case is obtained by switching $m_\phi$ and $m_\eta$), 
\begin{multline} \label{eq:class1-S}
    S = \frac{- n_2 d_3(\phi) a^2 v^2}{72\pi (m_\phi^2 - m_\eta^2)^2} \Bigg( 6 Y_\phi \left( 1 - \frac{m_\eta^2}{m_\phi^2} + \ln\frac{m_\eta^2}{m_\phi^2} \right) \\
    + \frac{5(m_\phi^6 - m_\eta^6) - 27 m_\phi^2 m_\eta^2(m_\phi^2-m_\eta^2) + 3(m_\phi^2 + m_\eta^2)(m_\phi^4 - 4 m_\phi^2 m_\eta^2 + m_\eta^4) \ln\frac{m_\eta^2}{m_\phi^2}}{(m_\phi^2 - m_\eta^2)^3} \Bigg) \,,
\end{multline}
\begin{equation}
    T = \frac{n_2 d_3(\phi) a^4 v^2}{192\pi^2 \alpha\, m_\phi^2 (m_\phi^2 - m_\eta^2)^5} \left( m_\phi^6 - m_\eta^6 + 9 m_\phi^2 m_\eta^2 ( m_\phi^2 - m_\eta^2) + 6 m_\phi^2 m_\eta^2 (m_\phi^2 + m_\eta^2) \ln\frac{m_\eta^2}{m_\phi^2} \right) \,.
\end{equation}

In Class~2, only the fermions contribute at one-loop. Assuming $d_2(\psi) > d_2(\chi)$ (the alternate case is obtained by switching $m_\psi$ and $m_\chi$),
\begin{multline}
    S = \frac{n_2 d_3(\psi) y_H^2 v^2}{36\pi (m_\psi^2 - m_\chi^2)} \Bigg( 6 Y_\psi \left( 2 + \frac{m_\psi^2 + m_\chi^2}{m_\psi^2 - m_\chi^2} \ln\frac{m_\chi^2}{m_\psi^2} \right) \\
    - \frac{2(m_\psi^8 + m_\psi^6 m_\chi^2 - m_\psi^2 m_\chi^6 - m_\chi^8) + 3(m_\psi^8 - 2m_\psi^6 m_\chi^2 + 6 m_\psi^4 m_\chi^4 - 2m_\psi^2 m_\chi^6 + m_\chi^8)\ln\frac{m_\chi^2}{m_\psi^2}}{(m_\psi^2 - m_\chi^2)^4} \Bigg) \,,
\end{multline}
\begin{multline}
    T = \frac{n_2 d_3(\psi) y_H^4 v^2}{192\pi^2 \alpha (m_\psi^2 - m_\chi^2)^5} \Bigg( 4 m_\psi^8 - 15 m_\psi^6 m_\chi^2 - 9 m_\psi^4 m_\chi^4 + 23 m_\psi^2 m_\chi^6  - 3 m_\chi^8 \\
    + 6 m_\psi^4 m_\chi^2 (m_\psi^2 - 5 m_\chi^2) \ln\frac{m_\chi^2}{m_\psi^2} \Bigg) \,.
\end{multline}

%==============================================================================
\section{\texorpdfstring{$Zf\bar{f}$ Couplings}{Zff Couplings}}
\label{app:Z-couplings}
%==============================================================================

The modifications to the $Zf\bar{f}$ couplings are described by the effective Lagrangian:
\begin{equation}
    \mathcal{L} = \frac{g}{\cos\theta_W} Z_\mu \bar{f} \gamma^\mu \left( (g_L + \delta g_L)P_L + (g_R + \delta g_R)P_R \right) f \,,
\end{equation}
where $f\in\{\tau,b\}$ and $g_{L,R}$ denote the momentum-dependent effective couplings in the SM (including radiative corrections) and $\delta g_{L,R}$ are the new physics contributions.  We choose to work in the on-shell scheme.

In Class~1 models the corrections are given by
\begin{align} 
    \delta g_L &= \frac{|y_\phi|^2 n_3}{128\pi^2} \sin^2(2\theta) \sum_{i=1}^{n_2} \left(g_{\eta_i} - g_{\phi_i} \right) F^{(1)}_Z(x_1,x_2) + \mathcal{O}(m_Z^2/M^2) \,, \label{eq:deltagL1} \\
    \delta g_R &= - \frac{|y_\eta|^2 n_3}{128\pi^2} \sin^2(2\theta) \sum_{i=1}^{n_2} \left(g_{\eta_i} - g_{\phi_i} \right) F^{(1)}_Z(x_1,x_2) + \mathcal{O}(m_Z^2/M^2) \,, \label{eq:deltagR1}
\end{align}
where $g_i = t^3_i - Q_i \sin^2\theta_W$. The loop function is 
\begin{multline}
    F^{(1)}_Z\left(x_1, x_2\right) = \frac{1}{x_1 - x_2} \Bigg( \frac{\left(x_2^2 + x_1(x_2 - 2)\right) x_2 \log x_2}{(x_2 - 1)^2} - \frac{\left(x_1^2 + x_2(x_1 - 2)\right) x_1 \log x_1}{(x_1 - 1)^2} \Bigg) \\
    + 2 + \frac{1}{x_1 - 1} + \frac{1}{x_2 - 1} \,.
\end{multline}
Note that although \cref{eq:deltagL1,eq:deltagR1} vanish in the limit of zero mixing, the sub-leading terms in $m_Z^2/M^2$ survive in this limit. Our numerical results use the full expressions. 

The expressions are more complicated in Class~2 models due to the fermion mixing. Here, the corrections take the form
\begin{align}
    \delta g_L &= \frac{|y_\psi|^2 n_3}{32\pi^2} \sum_{i=1}^{n_2} \left(g_{\psi_i} - g_{\chi_i} \right) F^{(2L)}_Z(x_1,x_2) + \mathcal{O}(m_Z^2/M^2) \,, \\
    \delta g_R &= - \frac{|y_\chi|^2 n_3}{32\pi^2} \sum_{i=1}^{n_2} \left(g_{\psi_i} - g_{\chi_i} \right) F^{(2R)}_Z(x_2,x_1) + \mathcal{O}(m_Z^2/M^2) \,,
\end{align}
with
\begin{multline}
    F^{(2L)}_Z\left(x_1, x_2\right) = \frac{1}{(x_1 - 1)(x_2 - 1)} \bigg( c_{\theta_R}^2 (c_{\theta_R}^2 - 2 c_{\theta_L}^2) (x_1 + x_2 - 2 x_1 x_2) + c_{\theta_R}^2 (x_1 - x_2) - 2 c_{\theta_L}^2 x_2 (x_1 - 1) \bigg) \\
    + \frac{1}{x_1 - x_2} \Bigg[ \frac{s_{\theta_R}^2 x_2 \log x_2}{(x_2 - 1)^2} \left( 2c_{\theta_L}^2 \Big(x_1 - x_2 - \frac{2x_2}{x_1} (x_2 - 1) \sqrt{x_1 x_2} \Big) + c_{\theta_R}^2 \left(x_2^2 + x_1(x_2 - 2)\right)\right) \\
    + \frac{c_{\theta_R}^2 x_1 \log x_1}{(x_1 - 1)^2} \left( - x_1 x_2 + 2 c_{\theta_L}^2 x_2 + c_{\theta_R}^2 x_2 (x_1 - 2) + s_{\theta_L}^2 \Big(x_1(2 - x_1) + \frac{4x_1}{x_2} (x_1 - 1) \sqrt{x_1 x_2}\Big) \right) \Bigg] \,,
\end{multline}
and $F^{(2R)}_Z(x_1,x_2)$ given by the same expression with $\theta_L \leftrightarrow \theta_R$.

The $Z$-boson couplings to fermions were precisely determined by the $Z$-pole measurements at LEP and SLC. The combined results for the $\tau$ couplings are~\cite{hep-ex/0509008}
\begin{equation}
    g_L^\tau = -0.26930 \pm 0.00058 \,, \qquad
    g_R^\tau = 0.23274 \pm 0.00062 \,,
\end{equation}
with a correlation of 0.44. The Standard Model prediction is $g_L^\tau = -0.26972$ and $g_R^\tau = 0.23262$~\cite{1902.05142}.

The $b$-quark forward-backward asymmetry $\mathcal{A}_{FB}^{0,b}$ measured at LEP-1 is in some tension with the global electroweak fit. Various groups have therefore performed fits to the data allowing for additional contributions to the $Z\bar{b}b$ vertex. We adopt the results of Ref.~\cite{1608.01509}:
\begin{equation}
    \delta g^b_L = 0.002 \pm 0.001 \,, \qquad
    \delta g^b_R = 0.016 \pm 0.006 \,,
\end{equation}
with a correlation of 0.90.

%==============================================================================
\section{Naturalness}
\label{app:naturalness}
%==============================================================================

All of the models introduce new states that couple to the Higgs and give radiative corrections to the Higgs mass at one-loop. This leads to a naturalness problem if these states have masses significantly greater than the weak scale. This can be quantified by considering the sensitivity of the Higgs mass in the IR to the "fundamental" parameters of the theory in the UV. Specifically, we adopt the fine-tuning measure\footnote{As discussed in \cite{1607.07446}, this fine-tuning measure can, within certain assumptions, be rigorously derived from a Bayesian model comparison.}
\begin{align} \label{eq:fine-tuning}
    \Delta \equiv \sqrt{ \sum_i \left(\frac{\alpha_i}{\mu_H^2} \frac{\partial \mu_H^2}{\partial \alpha_i}\right)^2 } \,,
\end{align}
where we approximate the Higgs mass parameter in the Standard Model Effective Field Theory (SMEFT) at the matching scale by $\mu_H^2 \approx (100\,\text{GeV})^2$, and $\alpha_i$ denote the parameters of the UV theory. We neglect the SMEFT running of $\mu_H^2$ between the matching scale and the weak scale, which has only an $\approx10\%$ effect on $\Delta$.

In Class~1 models only scalar diagrams give corrections to the Higgs mass at one-loop. Integrating these fields out and matching onto the SMEFT gives the matching condition:
\begin{equation}
    \mu_H^2 = \mu_{H,\text{UV}}^2 +\frac{a^2 n_2 d_3(\phi)}{16\pi^2} \left( 1 + \frac{m_\phi^2 \log(m_\phi^2/\mu_M^2) - m_\eta^2 \log(m_\eta^2/\mu_M^2)}{m_\eta^2 - m_\phi^2} \right)\,, 
\end{equation}
where $\mu_{H,\text{UV}}^2$ is the Higgs mass parameter in the UV theory. The couplings and masses appearing in the fine-tuning measure are  $\alpha_i \in \{\mu_{H,\text{UV}}^2, a, m_\phi^2, m_\eta^2\}$, where these are $\overline{\text{MS}}$ quantities evaluated at the matching scale\footnote{The scale at which one "defines" the UV theory is of course arbitrary. One could instead choose the $\alpha_i$ to be the masses and couplings at some high scale (for example $M_{Pl}$); this generally increases the fine-tuning.} $\mu_M$. This scale is taken to be geometric mean of the masses, $\mu_M=\sqrt{m_\phi m_\eta}$. (Note that the matching scale is fixed after evaluating the derivatives in \cref{eq:fine-tuning}.) The resulting expression for $\Delta$ is particularly simple in the limit of equal masses ($m_\phi=m_\eta$):
\begin{equation}
\label{eq:class1-fine-tuning}
    \Delta = \sqrt{1 + \frac{(a^2 n_2 d_3(\phi))^2}{512\pi^4\mu_H^4}} \approx \frac{1}{16\pi^2} \frac{a^2 n_2 d_3(\phi)}{\sqrt{2} \mu_H^2} \,,
\end{equation}
where the R.H.S. assumes $a \gg \mu_H$.

In Class~2, only the fermions contribute at one-loop and the matching condition is
\begin{equation}
    \mu_H^2 = \mu_{H,\text{UV}}^2 - \frac{y_H^2 n_2 d_3(\psi)}{8\pi^2} \left( m_\psi^2 + m_\chi^2 + \frac{m_\psi^4 \log(m_\psi^2/\mu_M^2) - m_\chi^4 \log(m_\chi^2/\mu_M^2)}{m_\chi^2 - m_\psi^2} \right) \,.
\end{equation}
In this case we have $\alpha_i \in \{\mu_{H,\text{UV}}^2, y_H, m_\psi, m_\chi\}$ and $\mu_M=\sqrt{m_\psi m_\chi}$. In the limit of equal masses ($M \equiv m_\psi = m_\chi$), the fine-tuning measure is
\begin{equation}
    \Delta = \sqrt{1 + \frac{y_H^2 n_2 d_3(\psi) M^2}{4\pi^2\mu_H^2} + \frac{7 (y_H^2 n_2 d_3(\psi))^2 M^4}{64\pi^4\mu_H^4}} \approx \frac{y_H^2 n_2 d_3(\psi)}{8\pi^2} \frac{\sqrt{7}M^2}{\mu_H^2} \,,
\end{equation}
where the approximation assumes $M \gg \mu_H$.

%==============================================================================
\section{Projected Sensitivities at Future Colliders}
\label{app:experimental-results}
%==============================================================================

With the LHC performing well, and the High Luminosity upgrade (HL-LHC) set to begin operations in 2027, there is currently a community-wide process aimed at understanding the benefits of various future collider possibilities.  This has led to various projections of future sensitivities under a range of assumptions on running scenarios.  Here we outline the assumptions underlying the sensitivity estimates we adopt.

%==============================================================
\begin{table}[ht]
    \begin{tabular}{@{\hspace{1em}} c @{\hspace{1.2em}} c @{\hspace{1.2em}} c @{\hspace{1.2em}} c @{\hspace{1.2em}} c @{\hspace{1.2em}} c @{\hspace{1.2em}} c @{\hspace{1.2em}} c @{\hspace{1em}}}
        \toprule
        &  HL-LHC & HE-LHC & ILC & CLIC & CEPC & FCCee & FCC \\
        \midrule
        $\Delta \kappa_\tau$   & 0.019 & 0.011 & 0.0070 & 0.0088 & 0.013 & 0.0073 & 0.0044\\
        $\Delta \kappa_b$      & 0.036 & 0.023 & 0.0058 & 0.0037 & 0.012 & 0.0067 & 0.0043\\
        $\Delta \kappa_\gamma$ & 0.019 & 0.012 & 0.034 & 0.022 & 0.037 & 0.039 & 0.0029\\
        $\Delta \kappa_g$     & 0.023 & 0.012 & 0.0097 & 0.009 & 0.015 & 0.010 & 0.0049\\
        \midrule
        $\Delta S$ & 0.081 & * & 0.079 & * & 0.014 & 0.0095 & *\\
        $\Delta T$ & 0.063 & * & 0.052 & * & 0.016 & 0.0068 & *\\
        \midrule
        $\Delta g_L^\tau$ & - & - & - & - & 0.00010 & 0.00002 & -\\
        $\Delta g_R^\tau$ & - & - & - & - & 0.00011 & 0.00002 & -\\
        $\Delta g_L^b$    & - & - & - & - & 0.00027 & 0.00030 & -\\
        $\Delta g_R^b$    & - & - & - & - & 0.00145 & 0.00162 & -\\
        \bottomrule
    \end{tabular}
    \caption{Projected sensitivities used in this analysis.  Projections for $\kappa_\tau$, $\kappa_b$, $\kappa_\gamma$ and $\kappa_g$ come from~\cite{deBlas:2019rxi}, those for $S$ and $T$ come from~\cite{1608.01509} (see text) and all others from~\cite{deBlas:2019wgy}.  
    The symbol "*" indicates no known reliable estimate while the symbol "-" indicates no significant improvement in precision.}
    \label{tab:experimental-results}
\end{table}
%============================================================================

For the Higgs measurements, the most up-to-date projected sensitivities at future colliders are the result of a global fit in the kappa framework~\cite{LHCHiggsCrossSectionWorkingGroup:2012nn,Heinemeyer:2013tqa}.  We take the projections for $\kappa_\tau$, $\kappa_b$ and $\kappa_\gamma$ from ~\cite{deBlas:2019rxi}, using the kappa-0 benchmark which assumes no new light particles which the Higgs boson can decay to.  Note that this means that the correlations between Higgs couplings present in our models will not be taken into account in the future projections.

While the running scenarios considered in~\cite{1608.01509} for $\Delta S$ and $\Delta T$ are not the most recent assumptions, the sensitivities quoted there are due to improvements in the measurements of the masses of the Higgs boson, the top quark and the W boson, which are unchanged from~\cite{1608.01509} to~\cite{deBlas:2019wgy}.  We assume that the current correlation between $S$ and $T$ is applicable to the future measurements.

Significant improvements in the $Z$ couplings are only expected with a $Z$-pole run at future electron-positron colliders.  We again assume that the current correlation between the measurements is applicable to the future measurements.

For our projected sensitivities we take the following assumptions~\cite{deBlas:2019rxi}:~HL-LHC collects 6.0 ab$^{-1}$ at 14\,TeV; for HE-LHC we take 15.0 ab$^{-1}$ at 27\,TeV under the S2' assumptions on theoretical uncertainties; for ILC$_{500}$ we take results assuming 2.0 ab$^{-1}$ at 250\,GeV, 0.2 ab$^{-1}$ at 350\,GeV and 4.0 ab$^{-1}$ at 500\,GeV; we consider the sensitivities of the entire CLIC programme, assuming 1.0 ab$^{-1}$ at 380\,GeV, 2.5 ab$^{-1}$ at 1.5\,TeV and 5.0 ab$^{-1}$ at 3.0\,TeV; for CEPC we take 16 ab$^{-1}$ at $m_Z$, 2.6 ab$^{-1}$ at $2m_W$ and 5.6 ab$^{-1}$ at 240\,GeV; for FCCee we again consider the entire programme, assuming 150 ab$^{-1}$ at $m_Z$, 10 ab$^{-1}$ at $2m_W$, 5 ab$^{-1}$ at 240\,GeV and 1.5 ab$^{-1}$ at $2m_t$; and for FCC we take the full FCC-ee/eh/hh programme, assuming FCCee given above along with 2.0 ab$^{-1}$ at 3.5\,TeV at FCC-eh and 30.0 ab$^{-1}$ at 100\,TeV at FCC-hh.

%=============================================================================
\bibliography{paper}
\bibliographystyle{JHEP}

\end{document}